\documentclass[onefignum,onetabnum]{siamart190516}
\usepackage{graphicx}
\usepackage{bookmark}
\usepackage[caption=false]{subfig} 
\graphicspath{{figures/}}
\definecolor{hotcolor}{rgb}{1,0,0}

\usepackage{lipsum}
\usepackage{amsfonts}
\usepackage{graphicx}
\usepackage{epstopdf}
\usepackage{algorithmic}
\ifpdf
  \DeclareGraphicsExtensions{.eps,.pdf,.png,.jpg}
\else
  \DeclareGraphicsExtensions{.eps}
\fi

\newsiamremark{remark}{Remark}
\newsiamremark{hypothesis}{Hypothesis}
\crefname{hypothesis}{Hypothesis}{Hypotheses}
\newsiamthm{claim}{Claim}

\headers{Reduced model of solute transport}{ R. MASRI, C. Puelz, and B. Rivi\`ere } 

\title{A reduced model for solute transport in compliant blood vessels with arbitrary axial velocity profile
}

\author{ Rami Masri\thanks{Department of Computational and Applied Mathematics, Rice University (\email{ rami.masri@rice.edu}, \email{riviere@rice.edu})} 
\and Charles Puelz \thanks{ Courant Institute of Mathematical Sciences, New York University (\email{puelz@cims.nyu.edu}).}
\and B\'eatrice Rivi\`ere  \footnotemark[2]}

\usepackage{amsopn}

\begin{document}
\maketitle 
\begin{abstract} 
We derive a reduced model of solute transport in blood based on the center manifold theory. 
The derivation is carried out on a convection diffusion equation with  general axial and radial velocity profiles in a blood vessel of varying cross section.
We couple the resulting one dimensional equation to a reduced model for blood flow in a compliant vessel. 
In the special case of a no--slip axial velocity profile, we study the dependence of the diffusion coefficient and corresponding numerical solutions on the shape of the profile. 

\end{abstract} 
\begin{keywords}
solute transport, reduced model, one dimensional blood flow, center manifold method  
\end{keywords}
\section{Introduction}

Modeling and computer simulation of solute transport in a fluid, like oxygen in blood, can provide critical insight on the planning of cardiovascular surgeries.  
These simulations remain computationally challenging due to the complexity of hemodynamics in vascular networks. 
Furthermore, the solute is transported by the blood, which is mathematically modeled by a coupled system of equations.
The numerical solution of the transport equation requires knowledge of the blood velocity field. 
The prohibitive cost of numerically solving the three dimensional Navier Stokes equations for the velocity field in large vessel networks has motivated the development of
reduced blood flow models as a computationally  efficient alternative \cite{mynard2015one,puelz2017computational}.

Our work derives a reduced model for the coupled flow and transport problem.  Flow in the blood vessel is approximated along its axial dimension by a one dimensional nonlinear hyperbolic system. The reduced model employs the radially averaged concentration of the solute, the cross--sectional area of the vessel
and the radially averaged velocity. The main contribution
of our work is a derivation of the reduced model in the general case where the radial component of the velocity is not neglected
and the shape of the axial velocity profile is not known in a vessel with impermeable boundary.  We employ the streamline boundary condition at the walls of the vessel, which is a condition also needed in the derivation of the reduced blood flow model \cite{barnard1966theory,vcanic2003mathematical}.
We note that the resulting one dimensional equation does not require  specification of a concentration profile. 
However, it is necessary to assume such a profile to provide closure to the equation derived using scaling and averaging arguments in \cite{d2007multiscale}.

Reduced models for blood flow without solute transport have been extensively studied in the literature. The reduced blood flow model is derived by performing an asymptotic analysis of the axisymmetric incompressible Navier Stokes equations \cite{barnard1966theory,vcanic2003mathematical}.    
This reduction yields a one dimensional nonlinear hyperbolic system that describes the dynamics of the vessel cross--sectional area and the blood velocity averaged over a cross section. 
The reduced model has been tested against measured physiological data and has produced similar qualitative and quantitative features \cite{olufsen2000numerical}. Further, this model is popular due to its low computational cost \cite{puelz2017comparison}.

For solute transport, several reduced models have been proposed under certain conditions. The reduction of a convection diffusion equation to a one dimensional model for the radially averaged concentration was first studied by Taylor  in the case of a pipe of constant radius and steady state Poiseuille flow  \cite{taylor1953dispersion,taylor1954conditions}.
Taylor assumed that diffusion renders the concentration of the solute radially uniform on a time scale that is faster than the convective transport.  Several mathematicians worked on providing a theoretical basis for this seminal paper \cite{taylor1953dispersion}. 
Aris calculated moments of the concentration using Fourier--Bessel analysis \cite{aris1956dispersion}.  We refer the reader to
the review of the different approaches to derive Taylor's equation, see \cite{young1991shear}. 
More recent work by Azer generalized Taylor's result to time dependent flow in a rigid pipe \cite{azer2005taylor}.
The center manifold theory provides another mathematical basis for Taylor's model.
The description of the center manifold theory is given in \cite{carr1983application} and its application to Taylor's dispersion problem was introduced by Mercer and Roberts for the case of Poiseuille flow \cite{mercer1990centre}.
The authors presented the method for the case of a straight channel and generalized it for a channel with varying radius based on a procedure given in \cite{roberts1988application}.
A numerical verification of the resulting equation for the case of laminar and turbulent flow was carried out in \cite{mohammed2014modelling}. 
Marbach and Alim recently used the generalized center manifold method to the case of a general velocity profile in a straight pipe, where the radial component of the velocity is neglected. 
Further, the authors derived a reduced model in a channel of varying radius for the specific case of Poiseuille flow  \cite{marbach2019active}.
In this work, we apply the center manifold method to a more general case where we do not assume a shape for the axial velocity profile, and we do not neglect the radial component of the velocity. 
To the best of our knowledge, such a derivation has not been carried before. 

Different choices for the axial velocity profile have been described in the literature. 
For example, Womersley theory was used to iteratively construct a time dependent velocity profile rather than specifying a steady state profile \cite{azer2007one}. 
Alternatively, a steady state profile may be specified.
Common choices are the no--slip and the flat velocity profiles. 
Puelz et al. performed a comparison of different no--slip profile shapes determined by the Coriolis coefficient, $\alpha$ \cite{puelz2017comparison}. 
The value of $\alpha$ appears in the viscous and convective terms of the blood flow model, and the authors concluded that it has a non--negligible impact on solutions. 
In our paper, for the specific case of a no--slip velocity profile, we specify a relation between the diffusion coefficient for the reduced transport equation and $\alpha$. 
Further, we perform numerical experiments to study the impact of different no--slip velocity profiles on the solute concentration.

The outline of this paper is as follows. In \Cref{sec: non-dim}, the transport problem and the center manifold method are described.
The main contribution of the paper is in \Cref{sec: model reduction}, where the reduced model  for a general axial and radial velocity profile is obtained. Particular cases are considered in \Cref{sec:particularcases}: flat velocity profile, no-slip velocity profile and Poiseuille flow. The reduced flow and transport model is applied to simulate momentum and concentration in a vessel in \Cref{sec: numerical experiments}. Conclusions are presented in \Cref{sec: conclusion}.

\section{Transport Model Problem and Center Manifold Method}
\label{sec: non-dim}

We consider an axisymmetric vessel and cylindrical coordinates $(x,r,\theta)$, where $x$ denotes the direction along the axis of symmetry of the vessel. We let $R(x,t)$ denote the inner vessel radius. 
Let $(V_x,V_r,V_{\theta})$ denote the velocity field of blood, where $V_\theta$ is assumed to be zero. We have the following equation modeling the concentration of a solute, $c(x,r,t)$, 
\begin{equation}
\frac{\partial c}{\partial t} + V_x\frac{\partial c}{\partial x}+V_r\frac{\partial c}{\partial r}  = D \left(\frac{\partial^2 c}{\partial r^2} + \frac{1}{r} \frac{\partial c}{\partial r} + \frac{\partial^2 c}{\partial x^2 } \right),
\label{eq:transport_cylin} 
\end{equation}
where $D$ is a constant diffusion coefficient.
We introduce the following characteristic quantities: inner vessel radius $R_0$,  length $\lambda$, concentration $c_0$, axial  velocity $U_0$ and radial velocity $V_0$. The non-dimensional variables are defined such that \cite{vcanic2003mathematical}:
\begin{equation}
r=R_0\bar{r}, \,\,  x= \lambda \bar{x}, \,\, t=\frac{\lambda}{V_0}\bar{t},\,\,  c=c_0\bar{c}, \,\, V_x=V_0\bar{V}_x,\,\, V_r= U_0\bar{V}_r,
\label{eq:non dimensional}
\end{equation}
where the following holds \begin{equation}
\frac{U_0}{V_0}  = \frac{R_0}{\lambda} = \epsilon_0.
\label{eq:epsilon}
\end{equation} 
Since the vessel's length is assumed to be much larger than its radius,  $\epsilon_0 \ll 1 $. This assumption is required in the derivation of the reduced blood flow model from the axially symmetric incompressible Navier Stokes equations in \cite{vcanic2003mathematical}.  

By substituting \cref{eq:non dimensional} in \cref{eq:transport_cylin}, the non-dimensional transport equation for the concentration $\bar{c}$ reads
\begin{multline} \frac{V_0}{\lambda}\frac{\partial (c_0\bar{c})}{\partial \bar{t}} + \frac{V_0}{\lambda}\bar{V}_x\frac{\partial (c_0 \bar{c})}{\partial \bar{x}}+ \frac{U_0 }{R_0}\bar{V}_r\frac{\partial  (c_0 \bar{c})}{\partial \bar{r}}  = \\
D \left(\frac{1}{R_0^2 } \frac{\partial^2 (c_0\bar{c})}{\partial \bar{r}^2} +  \frac{1}{R_0^2 } \frac{1}{\bar{r}} \frac{\partial (c_0 \bar{c})}{\partial \bar{r}} + \frac{1}{\lambda^2} \frac{\partial^2 (c_0\bar{c})}{\partial \bar{x}^2} \right).
\label{eq:first subs}
\end{multline} 
Multiplying \cref{eq:first subs} by $\lambda / V_0$, 
and noting from \cref{eq:epsilon} that $U_0\lambda = V_0 R_0$, we obtain
\begin{align}
\frac{\partial (c_0\bar{c})}{\partial \bar{t}} + \bar{V}_x\frac{\partial (c_0\bar{c})}{\partial \bar{x}}+\bar{V}_r \frac{\partial (c_0\bar{c})}{\partial \bar{r}}  = D\frac{\lambda}{V_0R_0^2 }\left(\frac{\partial^2 (c_0\bar{c})}{\partial \bar{r}^2} +  \frac{1}{\bar{r}} \frac{\partial (c_0 \bar{c})}{\partial \bar{r}} + \frac{R_0^2}{\lambda^2} \frac{\partial^2 (c_0\bar{c})}{\partial \bar{x}^2} \right). 
\label{eq:third subs}
\end{align}
Neglecting the terms of order $\epsilon_0^2$, equation \cref{eq:third subs} in non-dimensional form is reduced to
\begin{equation}
\frac{\partial (c_0\bar{c})}{\partial \bar{t}} + \bar{V}_x\frac{\partial (c_0 \bar{c})}{\partial \bar{x}}+\bar{V}_r \frac{\partial  (c_0\bar{c})}{\partial \bar{r}}  = D\frac{\lambda}{V_0R_0^2 }\left(\frac{\partial^2 (c_0\bar{c})}{\partial \bar{r}^2} +  \frac{1}{\bar{r}} \frac{\partial (c_0 \bar{c})}{\partial \bar{r}}  \right). 
\label{eq:transport_cylin_nondim} 
\end{equation} 
We rewrite this equation in dimensional variables and obtain
\begin{equation}
    \frac{\partial c}{\partial t} + V_x\frac{\partial c}{\partial x}+V_r\frac{\partial c}{\partial r}  = \mathcal{L}c, 
    \label{eq:equation we are solving}
    \end{equation}
where $\mathcal{L} $ is the following operator: 
    \begin{equation}
    \mathcal{L}c = D \left(\frac{\partial^2 c}{\partial r^2} + \frac{1}{r} \frac{\partial c}{\partial r} \right).
    \end{equation}
We note that this scaling argument was also employed by D'Angelo, where averaging arguments and assumptions on the concentration profile were subsequently used \cite{d2007multiscale}. We will use the center manifold method to arrive at a reduced model of \cref{eq:equation we are solving} without a priori assuming a profile for the concentration: a relation between $c$ and its radial average. This model will depend on the area of the vessel and the radially averaged quantities of $c$ and $V_x$.   
In order to carry out the derivation, we make several assumptions. First, we assume that the wall of the vessel is impermeable. This condition is also used in \cite{mercer1990centre,taylor1953dispersion,marbach2019active}, and reads  
\begin{equation}
    \frac{\partial c}{\partial r}\bigg\vert_{r=R(x,t)} = 0. 
    \label{eq:impermeability condition}
\end{equation}
We also assume blood is an incompressible fluid. This condition in cylindrical coordinates reads
\begin{equation}
\frac{\partial (rV_r)}{\partial r} + r \frac{\partial V_x }{ \partial x } = 0.
\label{eq:incompressibility condition}
\end{equation} 
Further, we assume the streamline boundary condition: 
\begin{equation} 
\frac{\partial R}{ \partial t } + V_x|_{r=R} \frac{\partial R}{\partial x } = V_r|_{r=R}.
\label{eq:streamline condition}
\end{equation}
Conditions \cref{eq:incompressibility condition} and \cref{eq:streamline condition} are also essential in the derivation of the reduced blood flow model \cite{vcanic2003mathematical}. 
\subsection{Application of the Center Manifold Method} 
\label{sec: method description} 
This section presents the center manifold theory applied to \cref{eq:equation we are solving} in the general case where
the radial velocity, $V_r$, is not neglected.
Let the partial Fourier transform of a function $f$ be denoted by  $\hat{f}$:
\begin{equation}
    \hat{f}(k,r,t) = \int_{-\infty}^{\infty} e^{ikx} f(x,r,t) dx. 
\end{equation} 
 Taking the Fourier transform of \cref{eq:equation we are solving}, we obtain 
\begin{equation}
\frac{\partial \hat{c}}{\partial t } + \hat{V_x} * \left( - ik  \hat{c} \right) +  \hat{V_r} * \frac{\partial \hat{c}}{\partial r} = \mathcal{L}\hat{c}.
\label{eq:fourier transform}
\end{equation}
Following \cite{mercer1990centre}, we supplement \cref{eq:fourier transform} with 
\begin{equation}
\frac{\partial k }{\partial t} = 0. 
\label{eq:wavenumber}
\end{equation}
The strategy of adding equation \eqref{eq:wavenumber} is similar to the approach used in dynamical systems \cite{wiggins2003introduction}. Mohammed et al. and Roberts argue that this equation represents the physical assumption that the concentration is slowly varying along the channel after a certain period of time \cite{mohammed2014modelling,roberts1997low}.
 This equation allows us to view \eqref{eq:fourier transform} and \eqref{eq:wavenumber} as a dynamical system in the variables $(k, \hat{c})$ with a stationary point at $(0,0)$. Hence, the center manifold method can be applied \cite{mercer1990centre}. More specifically, we can write equations \eqref{eq:fourier transform}  and \eqref{eq:wavenumber} as 
\begin{equation}
\frac{\partial }{\partial t } \mathbf{u}  = A \mathbf{u} + F(\mathbf{u}),
\label{eq:matrix form}
\end{equation}
where 
\begin{equation} 
\mathbf{u} =  \begin{pmatrix} k \\ \hat{c} \end{pmatrix}, \quad A = \begin{pmatrix} 0 & 0 \\ 0 & \mathcal{L} \end{pmatrix}, \quad F(\mathbf{u})= \begin{pmatrix} 0 \\  - \hat{V_x} * \frac{\partial \hat{c}}{\partial x} -  \hat{V_r} * \frac{\partial \hat{c}}{\partial r}  
\end{pmatrix}.
\end{equation}
The operator $A$ has two zero eigenvalues and all the other eigenvalues are negative. An application of Theorem 1 from \cite{carr1983application} implies the existence of a center manifold $S$. The following ansatz is chosen:
\begin{equation}
\hat{c}(k,r,t) = \hat{W}(k,r,t; \langle \hat{c} \rangle),
\label{eq:center}
\end{equation}
where $\langle \hat{c} \rangle$ is the radial average of $\hat{c}$.  In the remainder of the paper, we will denote
by $\langle f\rangle$ the radial average of a function $f$ defined by
\begin{equation}
    \langle f \rangle(x,t) = \frac{2}{R^2} \int_0^R  f(x,r,t) r dr.
    \label{eq:average}
\end{equation}
We take the inverse Fourier transform of \cref{eq:center} and assume  that it depends on the average of $\langle c \rangle $: 
\begin{equation}
    c(x,r,t) = W(x,r,t; \langle c  \rangle). 
    \label{eq:center inverse}
    \end{equation}
We substitute \cref{eq:center inverse} in \cref{eq:equation we are solving} and obtain the following equation:
 \begin{equation}
    \frac{\partial W}{\partial t} + \frac{\partial W}{\partial \langle c \rangle } \frac{\partial \langle c \rangle }{\partial t } + V_x  \frac{\partial W}{\partial x}  + V_r \frac{\partial W}{\partial r }  = 
    \mathcal{L}W,
    \label{eq:general equation in W G}
    \end{equation} 
where $W$ is to be determined. 
The following ansatz for the flow on the center manifold is also considered:  
\begin{equation}
\frac{\partial \langle c \rangle }{\partial t } = G(x,t; \langle c \rangle ).
\label{eq: flow on manifold}
\end{equation}
\Cref{eq:center} and \cref{eq: flow on manifold}  represent the ansatz employed in \cite{mercer1990centre,coullet1983amplitude}. 
If the zero solution of the equation describing the flow on $S$ is stable, then Theorem 2 from \cite{carr1983application} asserts that a solution $\mathbf{u}$ of \eqref{eq:matrix form} approaches a solution on the center manifold exponentially fast in time. 

The objective is to solve for $W$ and $G$ from \cref{eq:general equation in W G}. Thus, we consider the following expansions \cite{mercer1990centre}:
\begin{align}
     W &= \sum_{i=0}^{\infty} W_i \left[r, t ; \left(\frac{\partial^j \langle c \rangle }{\partial x^j }\right)_{j=0,...,i}\right],
    \label{eq:expression for centre manifold} \\ 
        \frac{\partial \langle c \rangle }{\partial t} & = \sum_{i=1}^{\infty} G_i \left[t ; \left(\frac{\partial^j \langle c \rangle}{\partial x^j }\right)_{j=0,...,i}\right].
        \label{eq:expression for the equation on the centre manifold}
\end{align}
We substitute \cref{eq:expression for centre manifold} and \cref{eq:expression for the equation on the centre manifold} in \cref{eq:general equation in W G} and decouple the equation in the following way \cite{mercer1990centre}:
\begin{align}
    \mathcal{L}W_0 & = 0, \label{eq:equation for W_0}\\ 
    \mathcal{L} W_1 & = \frac{\partial W_0}{\partial t}+ \frac{\partial W_0}{\partial \langle c \rangle }G_1 + V_x \frac{\partial W_0}{\partial x} + V_r \frac{\partial W_0}{\partial r}, \label{eq:equation for W_1}\\
        \mathcal{L}W_{n+1} &= \frac{\partial W_n}{\partial t} + \sum_{l=1}^{n+1} \sum_{p=0}^{n-l+1} \frac{\partial W_{n+1-l}}{\partial \langle c \rangle^p } \frac{\partial^p G_l }{\partial x^p } + V_x \frac{\partial W_{n}}{\partial x} + V_r \frac{\partial W_n }{\partial r}, \quad  n > 1
        \label{eq: general w n+1 as in roberts}
\end{align}
where $\langle c \rangle^p = \partial^p \langle c \rangle / \partial x^p $.
In order to find an explicit solution of equations \cref{eq:equation for W_0}, \cref{eq:equation for W_1} and \cref{eq: general w n+1 as in roberts}, we impose the following conditions \cite{marbach2019active,mercer1990centre}. 
\begin{align}
\frac{\partial W_i}{\partial r}|_{r = R(x,t)}   &= 0,
\label{eq:impermeability condition for W_i}  \\     
\langle W_0 \rangle & = \langle c \rangle, \label{eq:condition on W_0} \\ 
    \langle W_i \rangle & = 0, \quad i \geq 1. \label{eq:condition on W_i}
\end{align}
Equation \cref{eq:impermeability condition for W_i} ensures the impermeability condition of the solute at the wall of the vessel \cref{eq:impermeability condition} is satisfied. Conditions \cref{eq:condition on W_0} and \cref{eq:condition on W_i} ensure consistency in the sense that $\langle c \rangle = \langle W \rangle$.

\section{Model Reduction}
\label{sec: model reduction}

This section contains our main result, namely a derivation of the reduced convection diffusion equation for the solute. The objective is to find an expression for $G_n$ from \cref{eq:equation for W_0}, \cref{eq:equation for W_1} and \cref{eq: general w n+1 as in roberts}. We first solve for $W_0$. We multiply \cref{eq:equation for W_0} by $r$, integrate with respect to $r$ and use conditions \cref{eq:impermeability condition for W_i} and \cref{eq:condition on W_0}. Thus, we obtain
\begin{equation}
W_0 = \langle c \rangle.  
\label{eq:value of W_0}
\end{equation}
 We substitute \cref{eq:value of W_0} in \cref{eq:equation for W_1} and \cref{eq: general w n+1 as in roberts} for $n=1$. This yields:
\begin{align} 
    \mathcal{L}W_1 & = G_1 + V_x \frac{\partial \langle c \rangle }{\partial x},
    \label{eq: W_1 1} \\ 
    \mathcal{L}W_2 & = \frac{\partial W_1}{\partial t} + G_2 + 
\frac{\partial W_1}{\partial \langle c\rangle} G_1 +
\frac{\partial W_1 }{\partial \langle c \rangle'} \frac{\partial G_1}{\partial x} + V_x \frac{\partial W_1}{\partial x} + V_r \frac{\partial W_1}{\partial r}. \label{eq: W_2 2 } 
\end{align} 
The notation $\langle c\rangle'$ is used for $\partial \langle c\rangle/\partial x$.
We multiply \cref{eq: W_1 1} by $r$ and integrate once with respect to $r$. 
\begin{align}
D  r\frac{\partial W_1 }{\partial r }  = \int_0^r G_1 sds  + \frac{\partial \langle c \rangle }{\partial x }\int_0^r V_x(x,s,t) sds  + K(x,t),
\label{eq:integrating once}
\end{align}
where $K(x,t)$ is a function independent of $r$. Under the assumption that $\partial W_1 / \partial r$ is bounded at $r=0$, we must have: 
\begin{equation}
    K(x,t) = 0.
    \label{eq:value of the constant}
\end{equation}
 Using the impermeability condition \cref{eq:impermeability condition for W_i}, we note that: 
\begin{equation}
 0 = D  R\frac{\partial W_1 }{\partial r }\bigg \vert_{r= R} = \frac{R^2}{2} \left( G_1 +  \frac{\partial \langle c \rangle }{\partial x } \langle V_x \rangle \right),  \quad \forall x,t 
\end{equation}
 We conclude that $G_1$ is the following: 
 \begin{equation}
 G_1 = -  \frac{\partial \langle c \rangle}{\partial x}\langle V_x \rangle.
\label{eq:value of G_1}
\end{equation}
We substitute \cref{eq:value of G_1} in \eqref{eq: W_1 1}, multiply by $r$ and integrate with respect to $r$ to obtain
 \begin{equation}
   D  r\frac{\partial W_1 }{\partial r }   = \frac{\partial \langle c \rangle}{\partial x } \int_0^r (V_x(x,s,t) - \langle V_x \rangle(x,t)) s ds.
 \end{equation}
We solve for $W_1$ by integrating over $r$ and using \cref{eq:condition on W_i}. We obtain
\begin{equation}
    W_1 = \frac{\partial \langle c \rangle}{\partial x } \tilde{\eta}, \quad \mbox{with} \quad
\tilde{\eta} =  \eta -\langle \eta\rangle,
\label{eq:value of W_1}
     \end{equation}
and $\eta$ is defined by
 \begin{equation}
     \eta(x,r,t)  = \frac{1}{D}\int_0^r \frac{1}{z} \int_0^z (V_x(x,s,t)-\langle V_x \rangle(x,t)) sdsdz.  \label{eq:def of eta} 
\end{equation}
We substitute  \cref{eq:value of W_1} and \cref{eq:value of G_1} in \cref{eq: W_2 2 } and obtain the following equation for $G_2$.
 \begin{align}
  \mathcal{L}W_2 & = \frac{\partial  }{\partial t }\left(\frac{\partial \langle c \rangle}{\partial x}\tilde{\eta} \right)+ G_2 - \tilde{\eta} \frac{\partial }{\partial x} \left( \frac{\partial \langle c \rangle}{\partial x} \langle V_x \rangle\right) + V_x\frac{\partial }{\partial x} \left(\frac{\partial \langle c \rangle}{\partial x}\tilde{\eta}\right) 
+ V_r \frac{\partial \langle c \rangle}{\partial x} \frac{\partial \tilde{\eta}}{\partial r}.
  \label{eq:after subs}
 \end{align} 
We multiply \cref{eq:after subs} by $r$ and write it in the following way 
\begin{multline}
  r \mathcal{L}W_2  =r\frac{\partial  }{\partial t }\left(\frac{\partial \langle c \rangle}{\partial x}\tilde{\eta} \right) +  r G_2 
- r\tilde{\eta} \frac{\partial }{\partial x} \left( \frac{\partial \langle c \rangle}{\partial x} \langle V_x \rangle \right) + r\frac{\partial}{\partial x} \left(\frac{\partial \langle c \rangle}{\partial x}  V_x \tilde{\eta} \right) \\ - r\tilde{\eta} \frac{\partial \langle c \rangle}{\partial x}\frac{\partial V_x}{\partial x}
+ r V_r \frac{\partial \langle c \rangle}{\partial x} \frac{\partial \tilde{\eta}}{\partial r}.  \label{eq:equation to solve for G_2}
 \end{multline}
Using  the incompressibility condition \cref{eq:incompressibility condition}, we have
\begin{align}
    r V_r \frac{\partial \tilde{\eta}}{\partial r} 
= \frac{\partial (r V_r \tilde{\eta})}{\partial r} 
-\tilde{\eta} \frac{\partial (r V_r)}{\partial r}
= \frac{\partial (r V_r \tilde{\eta})}{\partial r}
+ r \tilde{\eta} \frac{\partial V_x}{\partial x}.  \label{eq:last term}
\end{align}
Substituting \eqref{eq:last term} in \eqref{eq:equation to solve for G_2}, we obtain: 
\begin{equation}
     r \mathcal{L}W_2= r\frac{\partial  }{\partial t }\left(\frac{\partial \langle c \rangle}{\partial x}\tilde{\eta} \right)  
+  r G_2 - r\tilde{\eta} \frac{\partial }{\partial x} \left( \frac{\partial \langle c \rangle}{\partial x} \langle V_x \rangle  \right) + r\frac{\partial}{\partial x} \left(\frac{\partial \langle c \rangle}{\partial x}  V_x \tilde{\eta}  \right) 
+ \frac{\partial \langle c \rangle}{\partial x} \frac{\partial(r V_r \tilde{\eta})}{\partial r }.
      \label{eq:updated G_2}
    \end{equation}
We rewrite the third term in the right-hand side of the equation above by using the chain rule:
 \begin{align}
-\tilde{\eta} \frac{\partial }{\partial x} \left( \frac{\partial \langle c \rangle}{\partial x}\langle V_x \rangle \right) 
= -\frac{\partial }{\partial x} \left( \frac{\partial \langle c \rangle}{\partial x} \langle V_x \rangle \tilde{\eta} \right) + \frac{\partial \langle c \rangle }{\partial x} \frac{\partial \tilde{\eta}}{\partial x} \langle V_x \rangle.
\label{eq:chain rule to second term}
\end{align} 
With \cref{eq:chain rule to second term}, equation \cref{eq:updated G_2} becomes:
\[
r\mathcal{L}W_2 = \frac{\partial }{\partial t} \left( \frac{\partial \langle c \rangle}{\partial x} r  \tilde \eta \right)
+  r G_2 
+ \frac{\partial }{\partial x} \left(\frac{\partial \langle c \rangle}{\partial x} r \tilde{\eta} \tilde{V}_x\right) 
+ \frac{\partial \langle c \rangle }{\partial x} \frac{\partial (r V_r \tilde{\eta})}{\partial r }
+ r\frac{\partial \langle c \rangle }{\partial x}\frac{\partial \tilde{\eta}}{\partial x} \langle V_x \rangle, 
\]
where the function $\tilde{V}_x$ is defined by:
\[
\tilde{V}_x = V_x - \langle V_x \rangle.
\]
We integrate the equation above from $r=0$ to $r =R$. Due to the impermeability condition \cref{eq:impermeability condition for W_i}, 
the integral of $r \mathcal{L}W_2$ vanishes. 
Using Leibniz rule, we obtain:
\begin{align}
 0 = \frac{\partial }{\partial t} \left( \frac{\partial \langle c \rangle}{\partial x} \int_0^R  r \tilde \eta  dr  \right) 
- \frac{\partial \langle c \rangle }{\partial x} \frac{\partial R}{\partial t}R \tilde{\eta}|_{r=R} 
+  \frac{R^2}{2}G_2 + \frac{\partial }{\partial x} \left( \frac{\partial \langle c \rangle}{\partial x}  \int_0^R r\tilde{\eta}  \tilde{V}_xdr \right) & \nonumber\\   
- \frac{\partial \langle c \rangle }{\partial x} \frac{\partial R}{\partial x} R \tilde{\eta}|_{r=R} \tilde{V}_x|_{r=R}
+ \frac{\partial \langle c \rangle }{\partial x} R V_r|_{r=R} \tilde{\eta}|_{r=R}
 +\frac{\partial \langle c \rangle }{\partial x} \langle V_x \rangle \int_0^R  r\frac{\partial \tilde{\eta}}{\partial x}dr 
\label{eq:G2inter2}
\end{align}
By construction, the radial average of $\tilde{\eta}$ is zero, which means the first term above vanishes.
We also note that
\begin{equation}
    \int_0^R r \tilde{\eta} \tilde{V}_x dr =   
    \int_0^R r \eta \tilde{V}_x dr - \langle \eta \rangle \int_0^R r  \tilde{V}_x dr   
= \int_0^R r \eta \tilde{V}_x dr.
\label{eq:step0}
\end{equation}
Finally by integration by parts, the last term in \cref{eq:G2inter2} is rewritten as
\begin{align}
    \frac{\partial \langle c \rangle }{\partial x} \langle V_x \rangle  \int_0^R  r\frac{\partial \tilde{\eta}}{\partial x} dr & =\frac{\partial \langle c \rangle }{\partial x} \langle V_x \rangle  \left( \frac{\partial }{\partial x} \int_0^R r \tilde{\eta}dr 
- \frac{\partial R}{\partial x}R \tilde{\eta }|_{r=R}  \right) \nonumber  \\   
    & = -  \frac{\partial \langle c \rangle }{\partial x}\langle V_x \rangle\frac{\partial R}{\partial x}R \tilde{\eta}|_{r=R}.
\label{eq:step1}	
\end{align}
With \cref{eq:step0} and \cref{eq:step1}, equation~\cref{eq:G2inter2} becomes:
\[
    0 = \frac{R^2}{2}G_2 + \frac{\partial }{\partial x} \left( \frac{\partial \langle c \rangle}{\partial x} \int_0^R  r\eta \tilde{V}_xdr \right) 
- \frac{\partial \langle c \rangle }{\partial x} R \tilde{\eta}|_{r=R}\left( \frac{\partial R}{\partial t}  
+ V_x|_{r=R}\frac{\partial R}{\partial x} - V_r|_{r=R}  \right).
\]
 We use the streamline condition \cref{eq:streamline condition} and obtain the following expression for $G_2$. 
 \begin{equation}
 G_2 = - \frac{2}{R^2}    \frac{\partial }{\partial x} \left( \frac{\partial \langle c \rangle}{\partial x} \int_0^R  r \eta \tilde{V}_x dr \right)  = - \frac{1}{R^2} \frac{\partial }{\partial x} \left( \frac{\partial \langle c \rangle}{\partial x} \langle R^2 \eta \tilde{V}_x \rangle  \right).
\label{eq:G2final}
 \end{equation}
We seek a second order approximation to \cref{eq:expression for the equation on the centre manifold}. Thus, we justify the truncation of the series \eqref{eq:expression for centre manifold} and \eqref{eq:expression for the equation on the centre manifold} by the following lemma.
\begin{lemma}\label{lemma:claim on the nondimensionality}
    We can write $W_n$ and $G_n$ in the following way.
\begin{align}
W_n & =  \left(\frac{R_0^{2} V_0}{\lambda D}\right)^n c_0 \bar{W}_n, \quad  \hspace{1.5em} n \geq 0, \label{eq:nondim W} \\ 
G_n & = \left(\frac{R_0^{2} V_0}{\lambda D}\right)^{n-1} \frac{V_0 c_0}{\lambda}\bar{G_n}, \quad n \geq 1, \label{eq:nondim G}  
\end{align}
where $\bar{G}_n$ and $\bar{W}_n$ are functions in the non-dimensional variables and $ \lambda, V_0, R_0$ and $c_0$ are the characteristic variables given in \cref{eq:non dimensional}. If we assume that the ratio of diffusion time to advection time is small, we have
\begin{equation} 
    \frac{R_0^{2} V_0  }{\lambda D} = \mathcal{O}(\epsilon),  \quad \epsilon \ll 1 
    \label{eq:assumption to get convergence}
\end{equation}
then by neglecting the terms of order  $\mathcal{O}(\epsilon^n), n\geq 2$, expressions \eqref{eq:expression for centre manifold} and  \eqref{eq:expression for the equation on the centre manifold} are reduced to
\begin{equation}\label{eq:secondorderapprox}
c = W_0 + W_1, \quad
\frac{\partial \langle c \rangle}{\partial t }   = G_1 + G_2.
\end{equation}
\end{lemma}
For the sake of completeness, we prove \Cref{lemma:claim on the nondimensionality} in \Cref{sec: proof of lemma}.
Taylor derived assumption \cref{eq:assumption to get convergence} as a condition for the radial variation of concentration to decay much faster than its longitudinal convection \cite{taylor1953dispersion}.
This derivation was for the case of a rigid pipe and Poiseuille flow. 
Mercer and Roberts used this assumption to justify the truncation of the series in the case of a rigid pipe, with $V_r$ neglected and $V_x$ assumed to depend only on $r$ \cite{mercer1990centre}.
\Cref{eq:assumption to get convergence} was also used as an assumption by Azer in the case of a rigid pipe and $V_x$ assumed to depend on $r$ and time $t$ \cite{azer2005taylor}. Marbach and Alim also used this assumption to arrive at their reduced models \cite{marbach2019active}.

According to \cref{eq:secondorderapprox}, \cref{eq:value of G_1} and \cref{eq:G2final}, a 
second order approximation for equation \cref{eq:expression for the equation on the centre manifold} describing the averaged concentration of the solute $\langle c \rangle$ is: 
\begin{equation}
\frac{\partial \langle c \rangle }{\partial t } =  - \langle V_x \rangle \frac{\partial \langle c \rangle}{\partial x} - \frac{1}{R^2} \frac{\partial }{\partial x} \left(  \frac{\partial \langle c \rangle}{\partial x}\langle R^2 \eta \tilde{V}_x \rangle  \right).
\label{eq: avg of c}
\end{equation} 
Averaging the incompressibility condition \cref{eq:incompressibility condition} yields \cite{vcanic2003mathematical}:
\begin{equation}
    \frac{\partial R^2 }{\partial t} + \frac{\partial (R^2 \langle V_x \rangle)  }{ \partial x }  = 0. 
    \label{eq:incompressibility in averaged quantities}
\end{equation}
We multiply \cref{eq: avg of c} by $R^2$ and use \cref{eq:incompressibility in averaged quantities}. This leads to the following final form of the equation in averaged quantities: 
\begin{equation}
\frac{\partial (R^2 \langle c \rangle) }{\partial t} 
+ \frac{\partial (R^2 \langle V_x \rangle \langle c \rangle) }{ \partial x } 
+  \frac{\partial }{\partial x} \left( \frac{\partial \langle c \rangle}{\partial x} \langle R^2 \eta \tilde{V}_x \rangle   \right)  = 0.
\label{eq:final form before change of notation} 
\end{equation}
We are now ready to present the reduced flow and transport model.  For readability, we simplify notation and let $U$ and $C$ denote the averaged quantities:
 \begin{equation}
 U = \langle V_x \rangle , \quad C= \langle c \rangle.
 \end{equation}
Next, we define the scaled cross-sectional area, $A$, and the momentum, $Q$:   
 \begin{equation}
 A = R^2, \quad Q = A U.
 \end{equation}
The reduced model for the flow involves the unknowns $A$ and $Q$ \cite{vcanic2003mathematical,barnard1966theory}.  The coupled reduced flow and transport model is:\vspace{1ex}

\framebox{\parbox[c]{0.90\textwidth}{
\begin{align}
   \frac{\partial A}{\partial t} + \frac{\partial Q}{\partial x} & = 0, \\ 
\frac{\partial Q}{\partial t} + \frac{\partial}{\partial x}\left( \alpha \frac{Q^2}{A} \right) + \frac{A}{\rho} \frac{\partial p}{\partial x} &= 2\pi\nu R  \frac{\partial V_x}{\partial r}\bigg |_{r=R}, \\ 
\frac{\partial (A C) }{\partial t} + \frac{\partial (Q C) }{ \partial x } + \frac{\partial }{\partial x} \left(  \frac{\partial C }{\partial x}  \langle A\eta (V_x - \langle V_x \rangle)  \rangle \right) & = 0.
\label{eq:A-Q-C general V_x}
\end{align}
}}
\vspace{1ex}

We recall that $\eta$ is defined by \eqref{eq:def of eta}. The parameters in the blood flow model are the density $\rho$,
the kinematic viscosity $\nu$, and the Coriolis coefficient $\alpha = \alpha(x,t)$. The Coriolis coefficient is a correction parameter resulting from asymptotic analysis of the Navier--Stokes equations \cite{vcanic2003mathematical}. This parameter depends on the axial velocity $V_x$ as follows:
\begin{equation}
    \alpha(x,t) = \frac{2}{R^2U^2} \int_0^R V_x^2 r dr. 
    \label{eq:coriolis coefficient}
\end{equation}
To provide closure to the system, a state equation for the pressure $p$ and a profile for the axial velocity must be specified. 
The following section will further specialize the model by considering several velocity profiles.

\section{Particular Cases}
\label{sec:particularcases}
\subsection{Flat Velocity Profile}
\label{sec: flat velocity profile}
In the case of inviscid flow  ($\nu = 0$), a flat velocity profile may be assumed \cite{vcanic2003mathematical,puelz2017comparison}. This choice means the axial velocity is independent of $r$.  In other words, it is equal to its radial average.  Therefore, we have
\begin{equation}
V_x = \langle V_x \rangle = U.
\label{eq:flat velocity profile}
\end{equation}
which results in $\alpha = 1$. Clearly, this implies the reduced model for the transport equation is purely hyperbolic.  This model has been
derived in \cite{d2007multiscale}. For completeness,
we write the resulting reduced flow and transport model for the flat velocity profile.
\begin{align}
   \frac{\partial A}{\partial t} + \frac{\partial Q}{\partial x} & = 0,\label{eq:inviscid1}\\ 
\frac{\partial Q}{\partial t} + \frac{\partial}{\partial x} \left( \frac{Q^2}{A} \right) + \frac{A}{\rho} \frac{\partial p}{\partial x} &= 0,\label{eq:inviscid2}\\ 
\frac{\partial (A C) }{\partial t} + \frac{\partial (Q C) }{ \partial x }  & = 0.
\label{eq:inviscid3}
\end{align}

\subsection{No-Slip Velocity Profile}
\label{sec: No - Slip Velocity Profile}
\begin{figure}[h!]
    \centering
    \subfloat[Diffusivity constant $\beta$]{\label{subfig: diffusion}
    \includegraphics[scale=0.29]{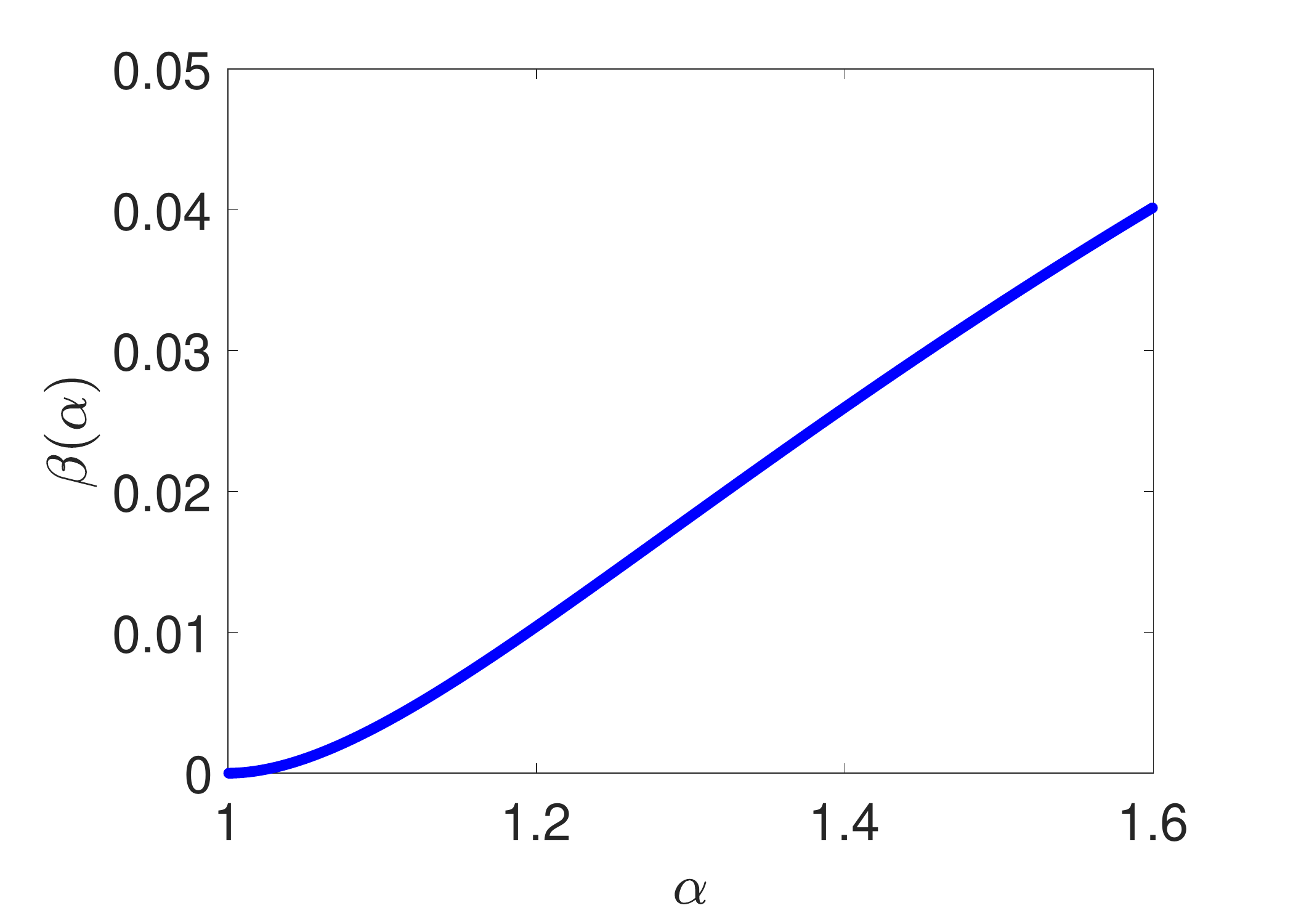}}
    \subfloat[Axial velocity profile $V_x$ ]{\label{subfig: different velocity profiles}\includegraphics[scale=0.29]{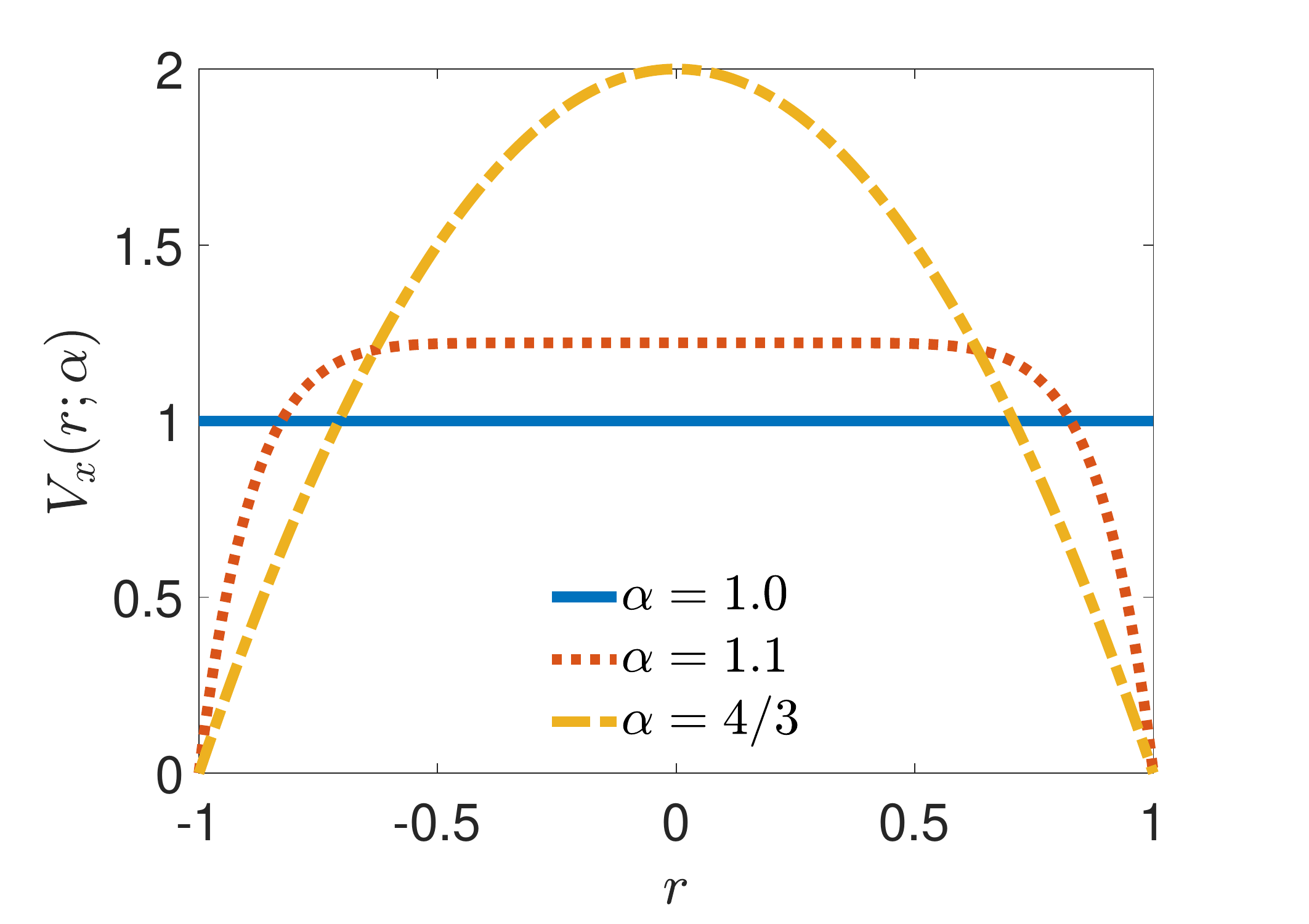}
    }
    \caption{Plot of the diffusivity constant $\beta$ and the velocity profile $V_x(r;\alpha)$ for different values of the Coriolis coefficient $\alpha$.}
    \label{fig: alpha vs conv and diff}
\end{figure}

The no--slip velocity profile is widely used for modeling blood flow.  This profile takes the form:
\begin{equation}
V_x(x,r,t) = \frac{\gamma + 2}{\gamma} U(x,t) \left(1-\left(\frac{r}{R}\right)^\gamma \right).
\label{eq:no slip velocity}
\end{equation}
We note that if $\gamma = 2$, \cref{eq:no slip velocity} reduces to the profile for Poiseuille flow. The case
$\gamma = 9$ has been shown to produce results that are good fits with experimental data \cite{vcanic2003mathematical,puelz2017comparison}.
The expression for $\tilde{V}_x$ can be explicitly defined:
\begin{equation}
\tilde{V}_x(x,r,t) = \frac{U(x,t)}{\gamma} \left( 2 -(\gamma + 2)\left(\frac{r}{R}\right)^\gamma \right).
\end{equation}
With this choice, we compute the expression for the function $\eta$:
\begin{equation}
\eta(x,r,t) = \frac{U(x,t)}{\gamma D} \left( \frac{r^2}{2} - \frac{r^{\gamma+2}}{(\gamma +2) R^\gamma}\right).
\end{equation}
We evaluate the diffusion coefficient in \eqref{eq:A-Q-C general V_x} when the no-slip velocity profile \eqref{eq:no slip velocity} is chosen to close the system. We find that 
\begin{equation}
A\langle \eta (V_x - \langle V_x \rangle)  \rangle = - \frac{\beta}{D} Q^2,
\end{equation}
where $\beta$ is a function of $\gamma$:
\begin{equation}
    \beta(\gamma) = \frac{1}{2(\gamma +2 )(\gamma + 4)}. 
\end{equation}
Substituting \cref{eq:no slip velocity} in \cref{eq:coriolis coefficient}, we obtain an expression for the Coriolis coefficient 
that depends on $\gamma$. In this case, $\alpha$ is independent of $x$ and $t$ and is given by:
\begin{equation}
\alpha = \frac{2+ \gamma}{\gamma + 1}.
\label{eq:gamma as a function of alpha}
\end{equation}
Equivalently, we have
\[
\gamma = \frac{2-\alpha}{\alpha - 1 }.
\]
The function $\beta$ is rewritten as a function of $\alpha$:
\begin{equation}
    \beta(\alpha) =  \frac{(\alpha -1)^2}{2\alpha (3\alpha - 2 )}.
    \label{eq: diffusion beta}
\end{equation}
We now state the reduced flow and transport model for the no--slip velocity profile.
\begin{align}
   \frac{\partial A}{\partial t} + \frac{\partial Q}{\partial x} & = 0, \label{eq:noslipeq1}\\
\frac{\partial Q}{\partial t} + \alpha \frac{\partial}{\partial x}\left( \frac{Q^2}{A} \right) + \frac{A}{\rho} \frac{\partial p}{\partial x} \label{eq:noslipeq2} &
= -2\pi\nu \frac{\alpha}{\alpha-1} \frac{Q}{A}, \\
\frac{\partial (A C) }{\partial t} + \frac{\partial (Q C) }{ \partial x } 
- \frac{(\alpha-1)^2}{2 \alpha (3\alpha -2) D} \frac{\partial }{\partial x} \left(  Q^2 \frac{\partial C }{\partial x} \right) & = 0.
\label{eq:noslipeq3}
\end{align}
We note that when $\alpha = 1$, $\beta = 0$. In this case, \cref{eq:noslipeq3} has no diffusion and we recover  equation  \cref{eq:inviscid3} for the flat velocity profile. In the remainder of the paper, the case ``$\alpha = 1 $'' refers to the flat velocity profile model \cref{eq:inviscid1}-\cref{eq:inviscid3}. Figure \ref{subfig: diffusion} shows the diffusion parameter,
$\beta$, as a function of the Coriolis coefficient $\alpha$ and Figure \ref{subfig: different velocity profiles} plots different shapes of the velocity profile, $V_x(r;\alpha)$, for several values of $\alpha$ when $U = 1$. 

\subsubsection{Poiseuille Flow} 
\label{subsec: Poiseuille Flow }
If  $\gamma = 2 $, or equivalently $\alpha = 4/3$, in \cref{eq:no slip velocity}, we have a Poiseuille flow given by 
\begin{equation}
V_x(x,r,t) = 2 U(x,t) \left( 1-\left(\frac{r}{R} \right)^2\right).
\label{v_x is Poiseuille Flow}
\end{equation} 
The value of the diffusivity constant \cref{eq: diffusion beta} in this case is $\beta = 1/48$,
which is the same diffusivity constant obtained by Taylor \cite{taylor1954conditions}. The reduced transport equation becomes:
\begin{equation}
    \frac{\partial (A C) }{\partial t} + \frac{\partial (Q C) }{ \partial x } - \frac{1}{48D} \frac{\partial }{\partial x} \left(  Q^2 \frac{\partial C }{\partial x}  \right)  = 0, 
    \label{eq:Poiseuille flow varying R}
\end{equation}
which is the same model derived in \cite{marbach2019active}. The authors in \cite{marbach2019active} use the center manifold method for the case when $V_x$ takes the form \eqref{v_x is Poiseuille Flow} and $V_r$ is derived from the incompressibility equation. 
If we further assume that the channel is of constant radius and that $U$ is constant, we recover  the equation derived  in \cite{taylor1954conditions,aris1956dispersion}:
\begin{equation}
\frac{\partial C}{\partial t} + U \frac{\partial C}{\partial x } -\frac{ U^2 A}{48 D} \frac{\partial^2 C }{\partial x^2}   = 0. 
\label{equation derived by Taylor}
\end{equation}
Finally, we remark that equations \eqref{equation derived by Taylor}  and  \eqref{eq:Poiseuille flow varying R}  agree with those derived in \cite{taylor1954conditions,marbach2019active} when the diffusion in the longitudinal direction is neglected in \cref{eq:transport_cylin}, see \eqref{eq:equation we are solving}. 

\section{Numerical Experiments}
\label{sec: numerical experiments}
In this section, we apply models \eqref{eq:inviscid1}--\eqref{eq:inviscid3} and  \eqref{eq:noslipeq1}--\eqref{eq:noslipeq3} to simulate flow and transport of a solute in a blood vessel.  To close the system, we choose the following equation of state for the pressure \cite{puelz2017comparison}
\begin{equation}
p(A)  = p_0 + \kappa (A^{1/2}-A_0^{1/2}), 
\label{state equation for pressure}
\end{equation}
where $A_0$ is a given reference area and $p_0$ is a given reference pressure.

\subsection{Sinusoidal pressure waveform}

\begin{figure}[H]
    \centering
   \subfloat{ \includegraphics[scale=0.29]{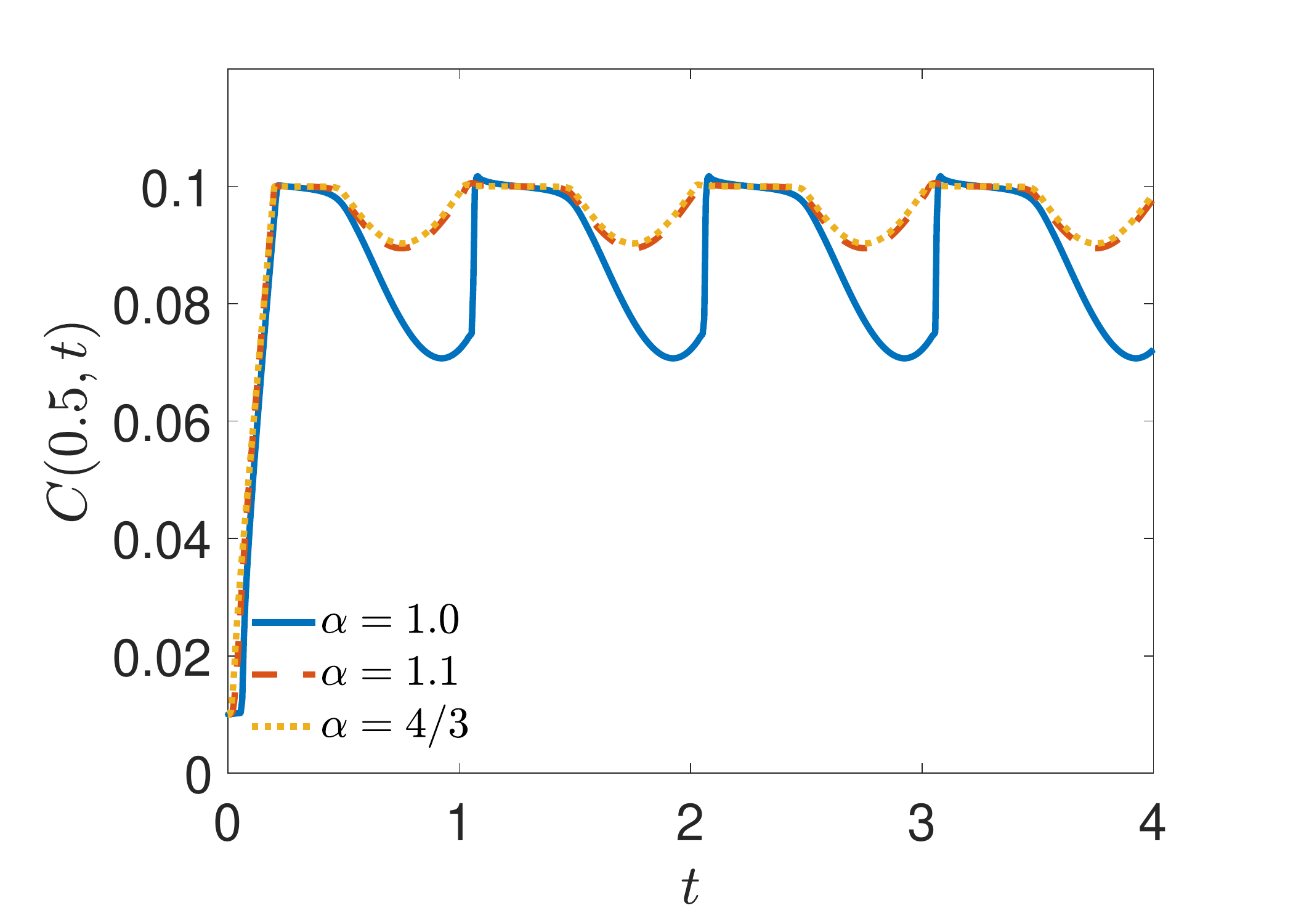} }
   \subfloat{\includegraphics[scale = 0.29]{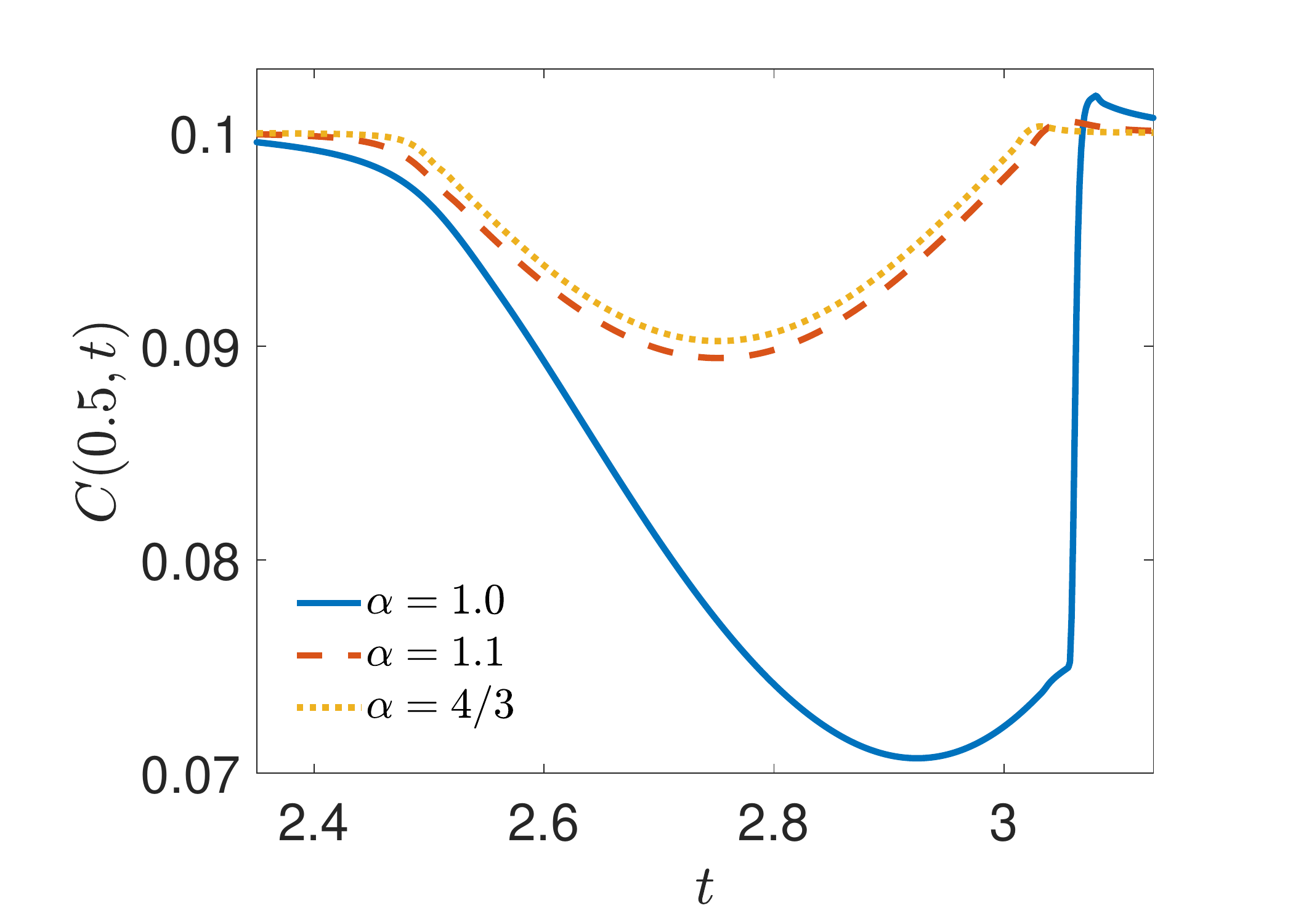}}
    \caption{Concentration, $C$, evaluated at the midpoint of the vessel as a function of time for different values of the Coriolis coefficient $\alpha$}
    \label{fig: concentration vs time sinusoidal for different alpha }
\end{figure}

\begin{figure}
\centering
    \subfloat{
    \includegraphics[scale=0.29]{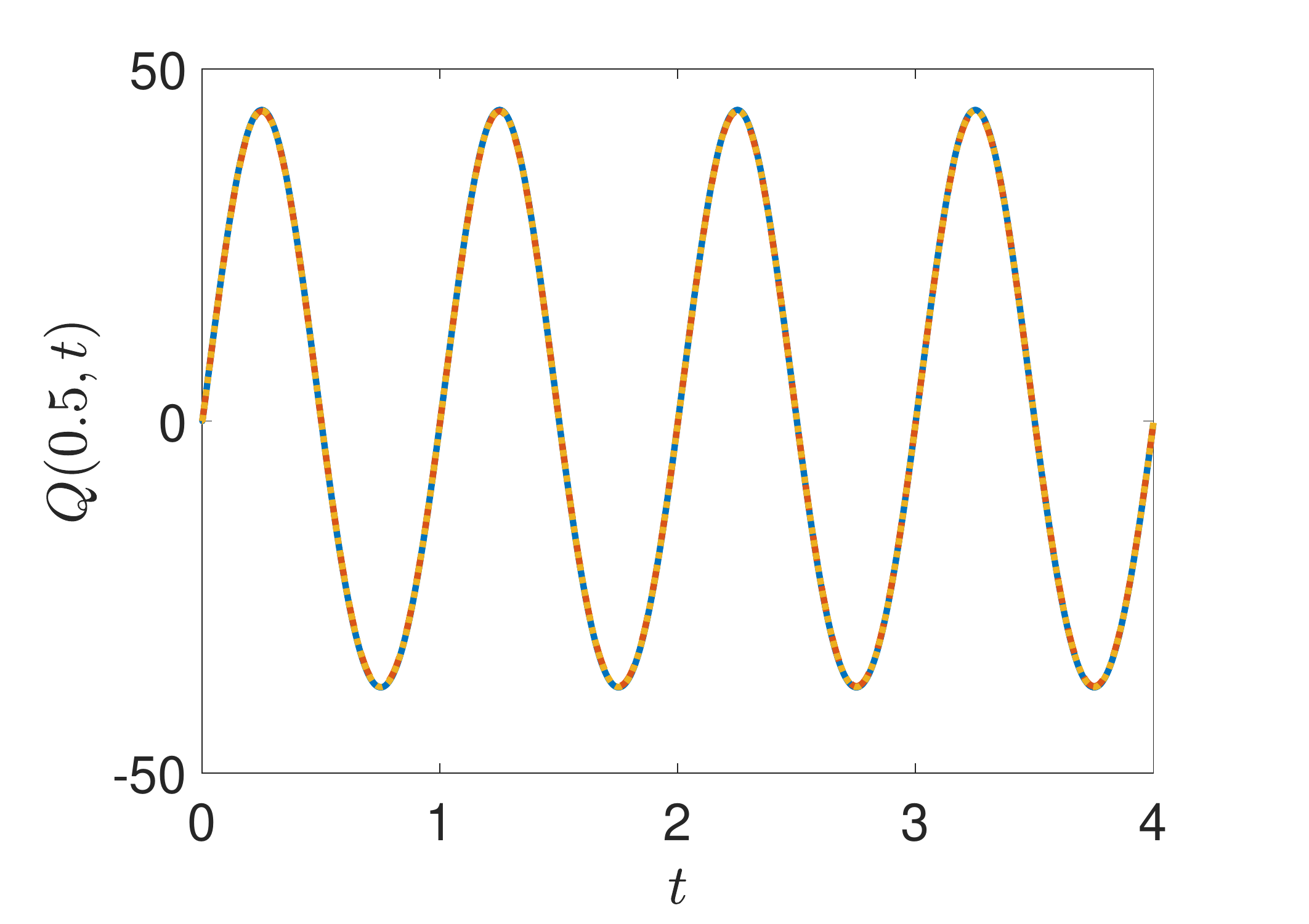}}
    \subfloat{
    \includegraphics[scale=0.29]{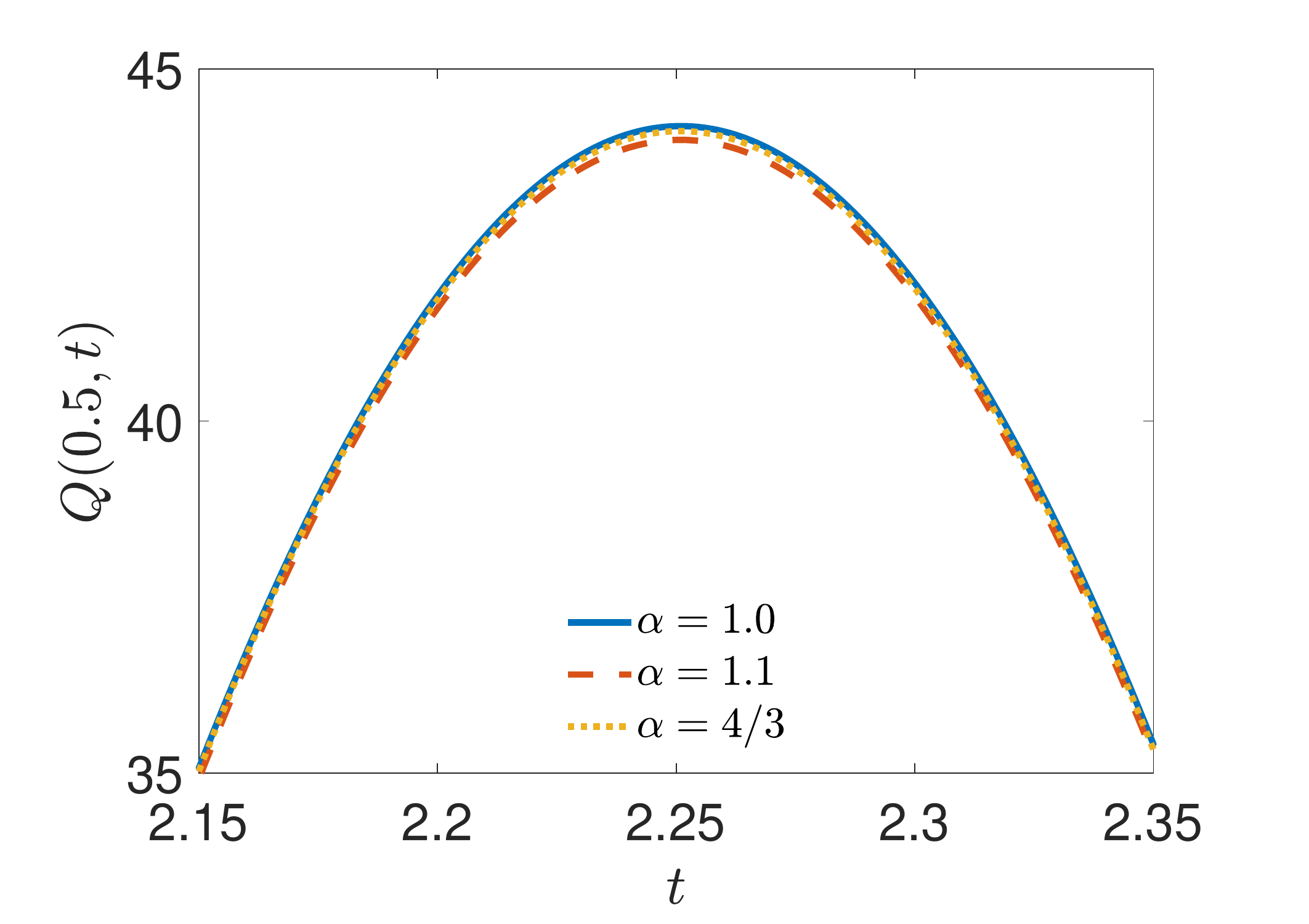}
    }
\caption{ Momentum, $Q$, evaluated at the midpoint of the vessel as a function of time for different values of the Coriolis coefficient $\alpha$}
\label{fig: momentum vs time sinusoidal for different alpha } 
\end{figure}

In the first numerical example, we impose a sinusoidal pressure waveform at the inlet of the vessel. 
\begin{equation}
    p(0,t) = 2\times 10^4 \, \sin(2 \pi t ) + p_0.
    \label{pressure}
\end{equation}    
We consider a vessel of length $L= 1 \text{cm}$ and set $\kappa = 4.5\times 10^5  \text{ g}/\text{s}^{2} \text{cm}^{2}$, $p_0 = 75$ mmHg, $A_0 = 1 \text{cm}$,  $\nu = 3.2 \times 10^{-2} \text{cm}^2 / \text{s}$ and the diffusion coefficient $D = 0.02 \text{cm}^2/\text{s}$. The inlet values for the area $A$ are determined from \eqref{state equation for pressure}. The momentum $Q$ at the inlet is specified by extrapolating the Riemann invariants of the $A$-$Q$ system \cite{puelz2017comparison}. We consider the following initial and boundary conditions for  \cref{eq:noslipeq3} when the direction of the flow is positive:
\begin{align}
    C(x,0) &= 0.01, \quad 0\leq x \leq L,  \\ \label{init C1}
    C(0,t) & = \begin{cases} 
    0.45 t + 0.01 & t \leq 0.2, \\ 
    0.1   & t > 0.2, 
    \end{cases} \\
    \frac{\partial C}{\partial x }(L,t) & = 0.  \label{inlet C}
\end{align} 
When the flow direction reverses, the value of the concentration at the outlet node is extrapolated and the inlet node ($x=0$) is treated like an outflow node. We numerically solve the $A$-$Q$ system, \eqref{eq:noslipeq1} and \eqref{eq:noslipeq2}, using the Runge Kutta discontinuous Galerkin scheme with the local Lax Friedrichs numerical flux \cite{puelz2017comparison}. After obtaining numerical solutions for $A$ and $Q$, we solve \cref{eq:noslipeq3} for $C$ using the non--symmetric interior penalty discontinuous Galerkin method and the local Lax Friedrichs numerical flux. 
The final simulation time is $T=4$ seconds. 
We choose three different values for $\alpha$ that have been used in the literature for blood flow: $1$ and $4/3$ are common values  and the value $1.1$ was shown to produce an accurate model when compared to experimental data \cite{vcanic2003mathematical,puelz2017comparison}. 
We recall that the model used for $\alpha = 1$ is \eqref{eq:inviscid1}-\eqref{eq:inviscid3}. For this particular case, we employ the min-mod slope limiter to further stabilize the discrete approximation of the concentration \cite{cockburn1989tvb}. 
\Cref{fig: concentration vs time sinusoidal for different alpha } and \cref{fig: momentum vs time sinusoidal for different alpha } show the evolution of the averaged concentration and momentum at the midpoint of the vessel for different values of $\alpha$.  We observe the momentum behaves as a sinusoidal function: when $Q$ is positive, the flow direction is from left to right and when $Q$ is negative
 the flow direction reverses from right to left.  The momentum profiles are nearly identical for the three values of the Coriolis coefficient. The situation is very different for the concentration profiles. We observe that the concentration profile dips much lower in the case
 of no diffusion ($\alpha = 1$) than in the case of Poiseuille flow ($\alpha = 4/3$) or the case $\alpha = 1.1$.  After a transition regime,
 all concentration profiles are periodic. We also observe that the change of flow direction has a direct impact on the value of the concentration as time evolves. 
\subsection{Flow and transport in ascending aorta with physiological boundary data}
\begin{figure}[H]
\centering
\subfloat{
\includegraphics[scale=0.29]{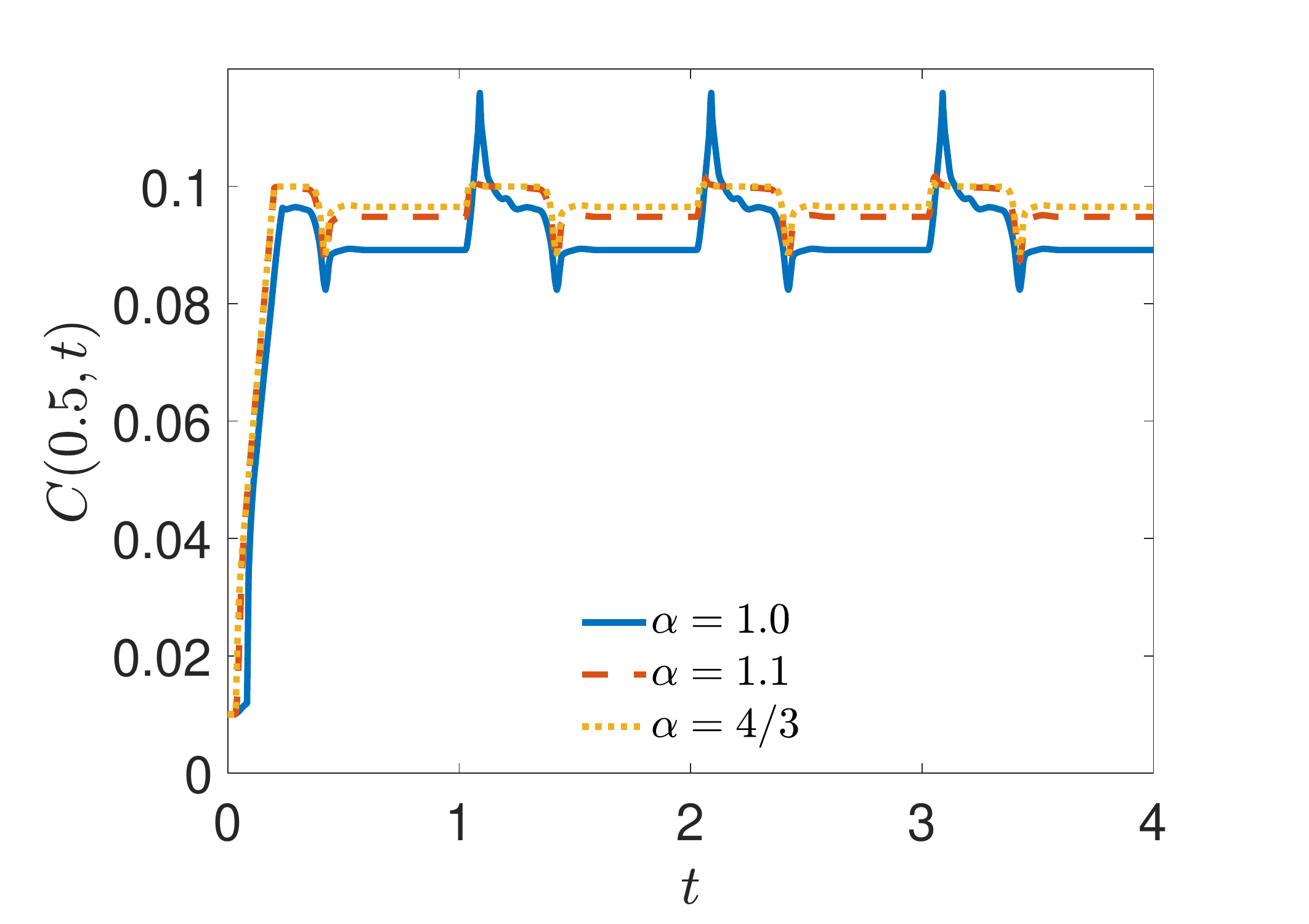}
} 
\subfloat{
\includegraphics[scale=0.29]{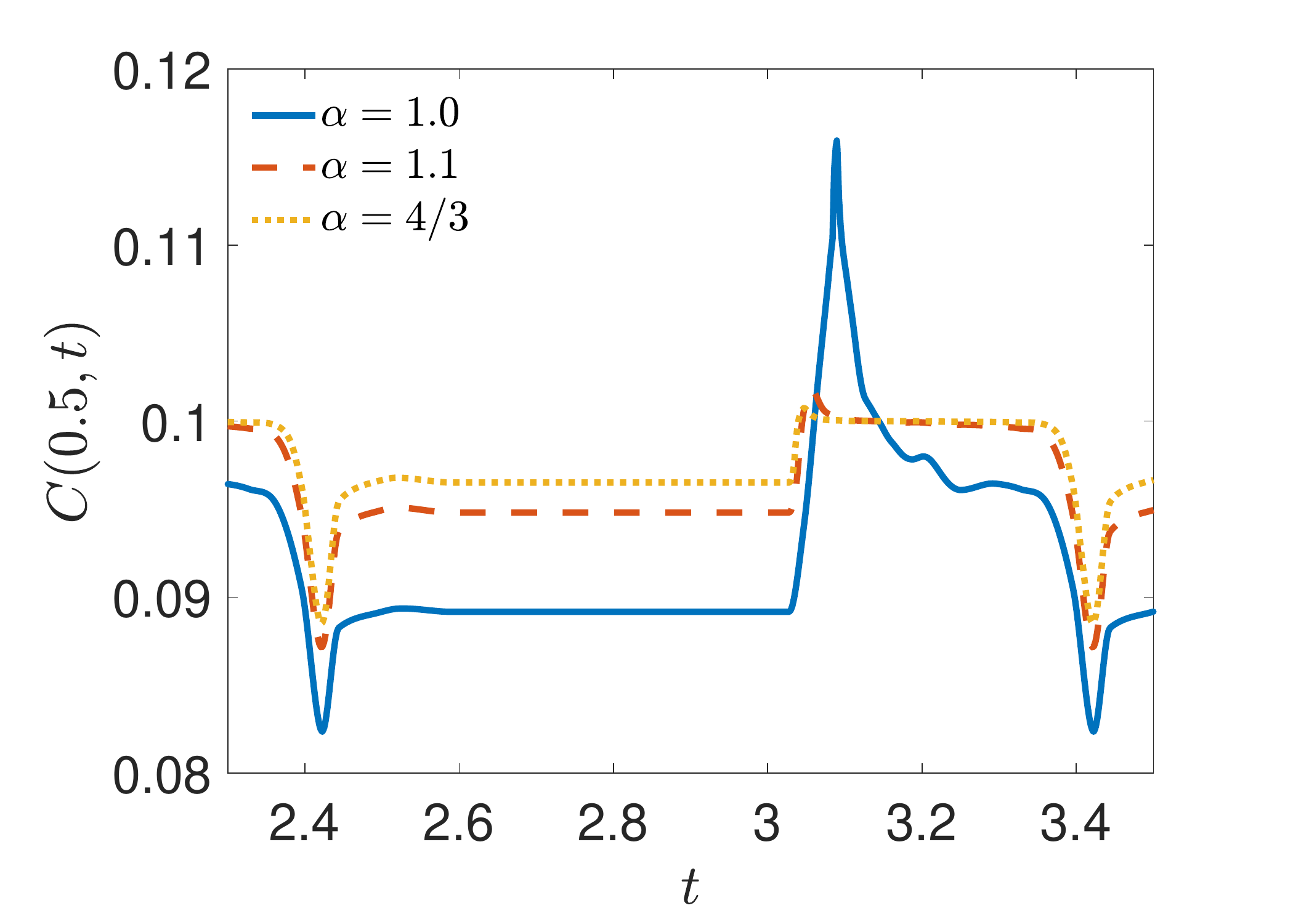} }
\caption{Concentration, $C$, evaluated at the midpoint of the ascending aorta as a function of time for different values of the Coriolis coefficient $\alpha$}
\label{fig: concentration vs time phys for different alpha } 
\end{figure}

\begin{figure}[H]
    \centering
    \subfloat{ 
    \includegraphics[scale=0.3]{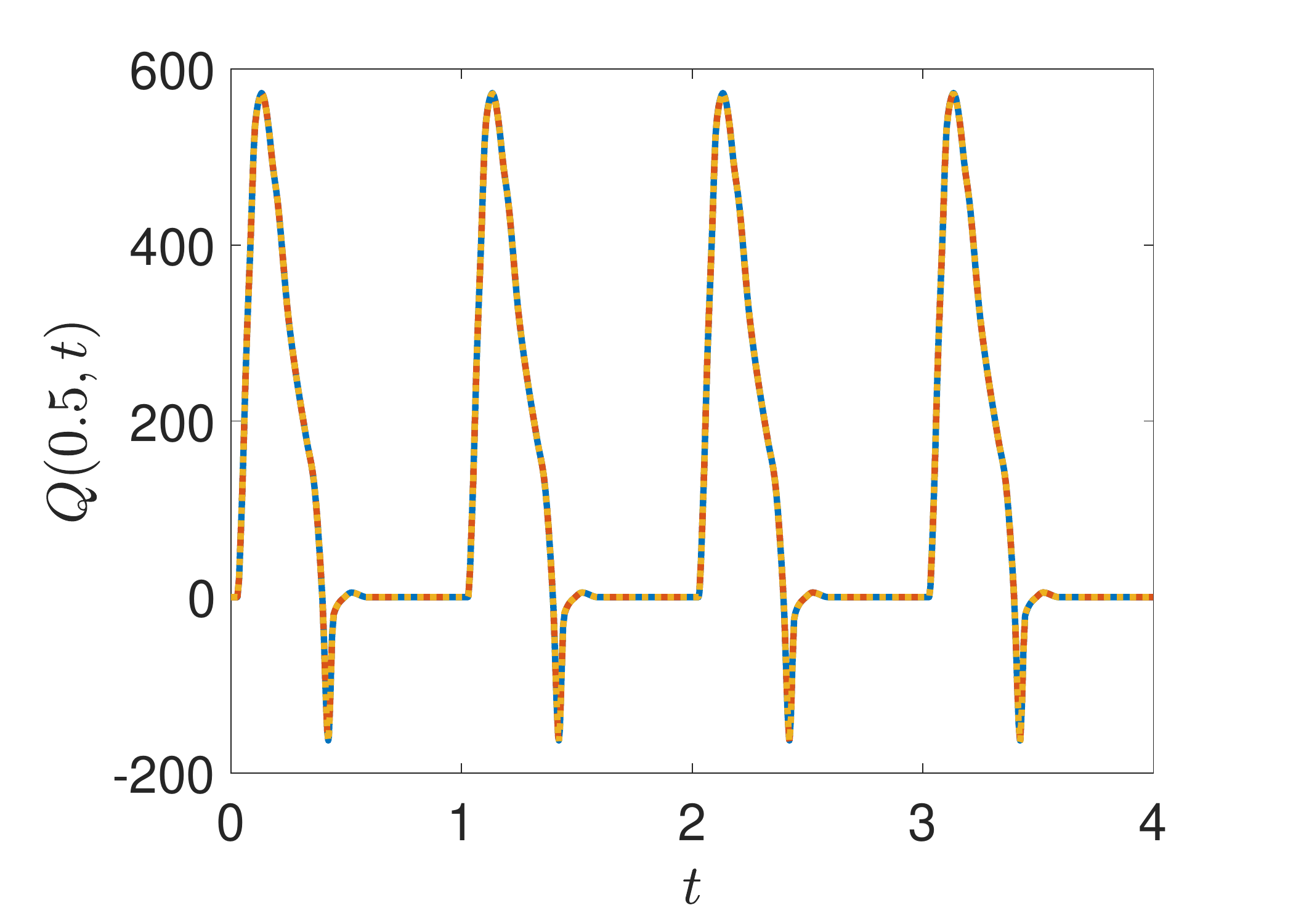}
    } 
    \caption{Momentum, $Q$, evaluated at the midpoint of the ascending aorta as a function of time for different values of the Coriolis coefficient $\alpha$}
    \label{fig: momentum vs time phys for different alpha }
\end{figure}

\begin{figure}[H]
\centering
\subfloat{\includegraphics [scale=0.29]{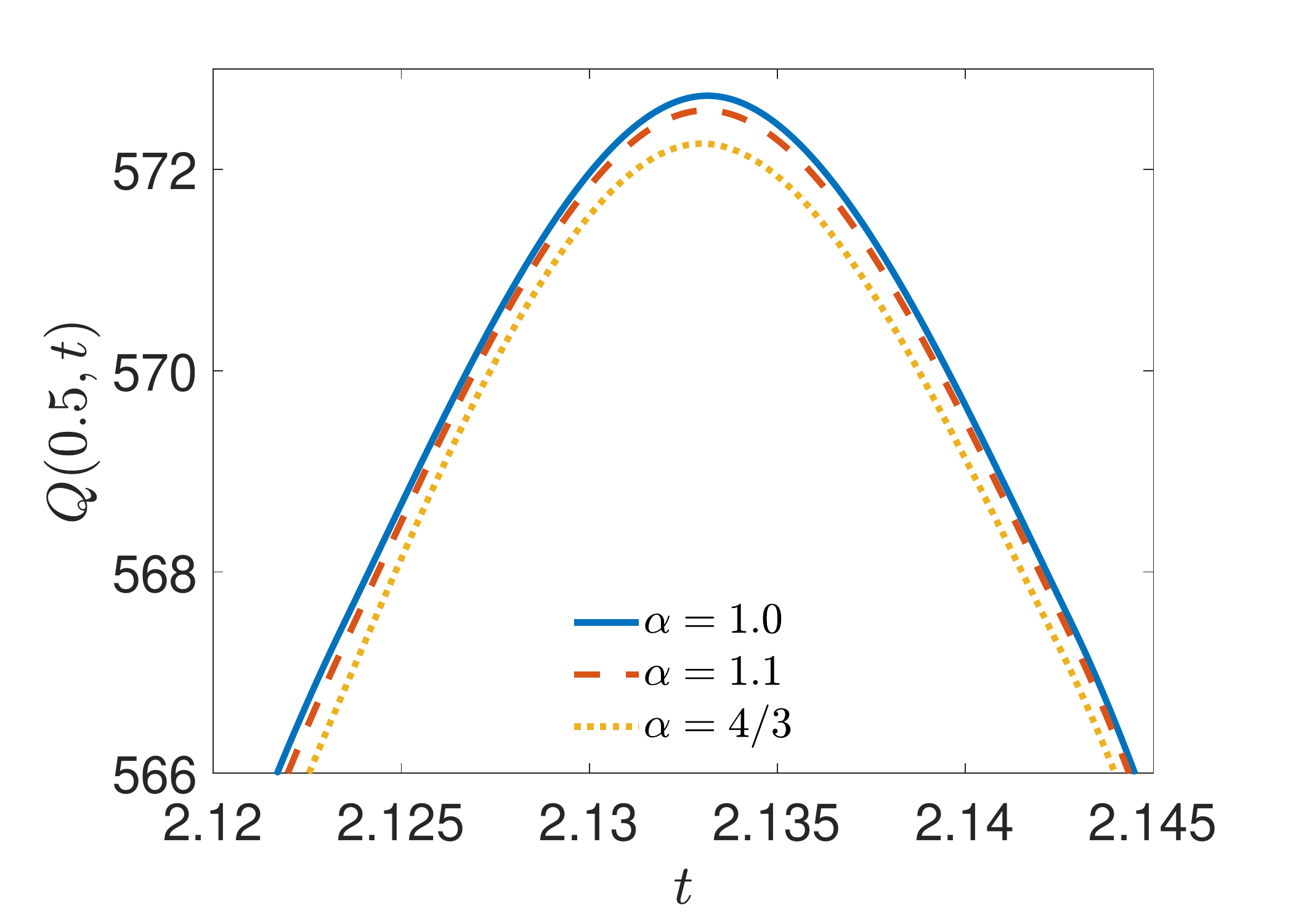}}
\subfloat{ \includegraphics[scale=0.29]{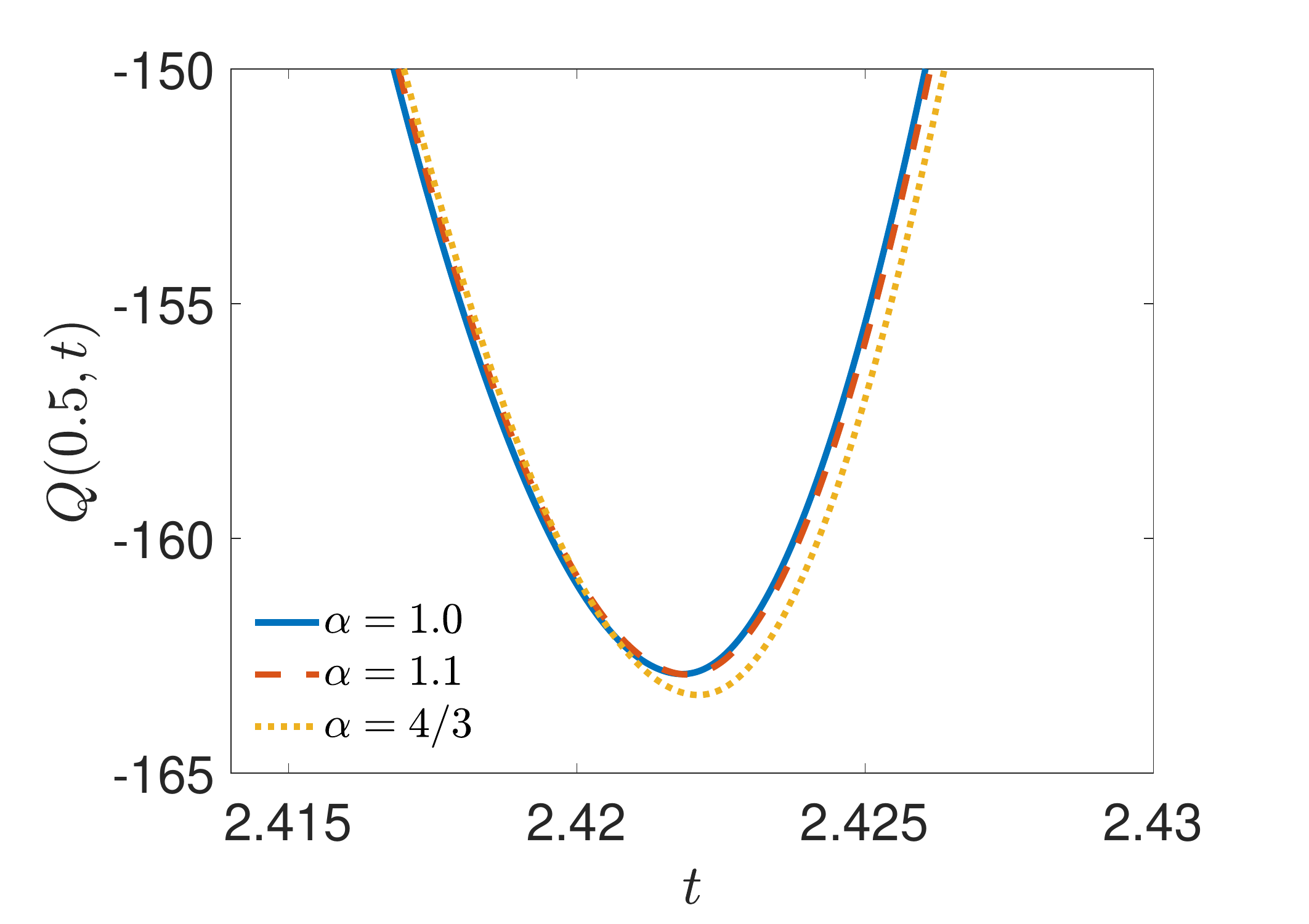}} 
\caption{Momentum, $Q$, evaluated at the midpoint of the ascending aorta as a function of time for different values of the Coriolis coefficient $\alpha$}
\label{fig: momentum vs time phys for different alpha zoomed} 
\end{figure}

This second example simulates flow and transport in the ascending aorta.  The input data for the reduced model is obtained from physiological data of the momentum, $Q$ \cite{boileau2015benchmark}. The physiological parameters are $ L = 4 \text{ cm}, A_0 = 5.983 \text{ cm}^2, p_0 = 75 \text{ mmHg},$ and $\kappa = 9.7\times 10^4  \text{ g}/\text{cm}^2$. We consider the same initial and boundary conditions for the concentration equation as in the previous example. \Cref{fig: concentration vs time phys for different alpha } and \Cref{fig: momentum vs time phys for different alpha } show the concentration, $C$, and the momentum, $Q$, evaluated at the midpoint of the ascending aorta. \Cref{fig: momentum vs time phys for different alpha zoomed} shows the maximum and minimum values of the momentum, $Q$, for the different values of $\alpha$. We observe that the momentum profiles  are periodic and attain different maximum values for the different Coriolis coefficients. The minimum values for  $\alpha = 1$ and $\alpha = 1.1$ are nearly identical whereas the Poiseuille case is shifted a little to the right. Similarly, after a transition period, the concentration profiles are periodic. We observe significant differences between the three concentration profiles.  These simulations show the significant impact of the different reduced models
on the solute concentration.

\section{Conclusion}
\label{sec: conclusion}
This paper contains a derivation of a reduced model for solute transport in a compliant vessel which allows for varying radius and arbitrary axial velocity profile. We recover well known models in the particular cases of Poiseuille flow or a flat velocity profile. We show that the Coriolis parameter has a significant
impact on the concentration profiles, in particular for problems with physiological data. Further work is needed to validate the various models with experimental data for blood flow and transport.

\section*{Acknowledgements} The authors thank Craig Rusin for the help with the computational results. 

\appendix \section{Justifying the Series Truncation}
\label{sec: proof of lemma}
In this section, we provide a proof by induction for \Cref{lemma:claim on the nondimensionality}.
\begin{proof}
We note that \cref{eq:nondim W}  holds for  $n=0$, $W_0 = \langle c \rangle  = c_0 \langle \bar{c} \rangle = c_0 \bar{W}_0$. We write the computed value of $G_1$, \cref{eq:value of G_1}, in non-dimensional form: \begin{align*} 
G_1 & = - \frac{\partial \langle c \rangle }{\partial x } \langle V_x \rangle 
=  \frac{V_0 c_0}{\lambda } \bar{G}_1,
\end{align*} 
where $\bar{G}_1 =  ( \partial \langle \bar{c} \rangle / \partial \bar{x} )\langle \bar{V}_x \rangle$. Thus, \cref{eq:nondim G} holds for $n=1$. Assume that \cref{eq:nondim W} and \eqref{eq:nondim G} hold for $ 1 \leq n \leq N $. We show that \cref{eq:nondim G} and \cref{eq:nondim W} hold for $n = N+1$. The equation for $W_{N+1}$ and $G_{N+1}$ is the following \cite{mercer1990centre}.    
\begin{align} 
\mathcal{L}W_{N+1} = \frac{\partial W_N}{\partial t} + G_{N+1} + \sum_{l=1}^{N} \sum_{p=0}^{N-l+1} \frac{\partial W_{N+1-l}}{\partial \langle c \rangle^p } \frac{\partial^p G_l }{\partial x^p } + V_x \frac{\partial W_{N}}{\partial x} + V_r \frac{\partial W_N }{\partial r}.
\label{general eq for w n+1 as in robers}
\end{align} 
We multiply \cref{general eq for w n+1 as in robers} by $r$, average it radially and multiply by $2/R^2$. We use the impermeability condition \cref{eq:impermeability condition for W_i} and obtain the following.  
\begin{align} 
0 &  = \left\langle \frac{\partial W_N}{\partial t} \right \rangle  + G_{N+1} +   \sum_{l=1}^{N} \sum_{p=0}^{N-l+1} \left\langle  \frac{\partial W_{N+1-l}}{\partial \langle c \rangle^p } \frac{\partial^p G_l }{\partial x^p } \right \rangle  +  \left \langle V_x \frac{\partial W_{N}}{\partial x} \right \rangle + \left \langle V_r \frac{\partial W_N }{\partial r} \right \rangle.
\label{after getting rid of W_n+1} 
\end{align} 
We note the following relation 
\begin{align*}
    \langle c \rangle^p = \frac{\partial^p \langle c \rangle }{\partial x^p} = \frac{c_0}{\lambda^p}\frac{\partial^p \langle \bar{c} \rangle }{\partial \bar{x}^p}= \frac{c_0}{\lambda^p}\langle \bar{c} \rangle^p. 
\end{align*}
Thus, we use the above equality,  \cref{eq:nondim W} for $n = N+1 - l$ and \cref{eq:nondim G} for $n=l$, where $l = 1,...,N$. We obtain: 
\begin{align}
\frac{\partial W_{N+1-l}}{\partial \langle c \rangle^p } \frac{\partial^p G_l }{\partial x^p }
& = \left( \frac{R_0^{2} V_0}{\lambda D}  \right)^{N} \frac{V_0c_0}{\lambda} \frac{\partial \bar{W}_{N+1-l}}{\partial \langle \bar{c}\rangle^p } \frac{\partial^p \bar{G}_l}{\partial \bar{x}^p}, \quad l = 1,...,N.
\end{align}
Then, \cref{after getting rid of W_n+1} in non-dimensional form reads:  
\begin{multline} 
0  = \left( \frac{R_0^{2} V_0}{\lambda D}  \right)^{N} \frac{V_0c_0}{\lambda} \left\langle \frac{\partial \bar{W}_N}{\partial \bar{t}} \right\rangle + \left(\frac{R_0^{2} V_0}{\lambda D}  \right)^{N} \frac{V_0c_0}{\lambda} \sum_{l=1}^{N} \sum_{p=0}^{N-l+1} \left \langle  \frac{\partial \bar{W}_{N+1-l}}{\partial \langle \bar{c}\rangle^p } \frac{\partial \bar{G}_l}{\partial \bar{x}} \right \rangle \\  + G_{N+1} + \left(\frac{R_0^{2} V_0}{\lambda D}  \right)^{N} \frac{V_0c_0}{\lambda} \left\langle \bar{V}_x \frac{\partial \bar{W}_N}{\partial \bar{x}}\right\rangle 
 +  \left( \frac{R_0^{2} V_0}{\lambda D}  \right)^{N}\frac{U_0 c_0}{R_0}\left \langle \bar{V}_r \frac{\partial \bar{W}_N}{\partial \bar{r} } \right \rangle \nonumber.
\end{multline}
We note that by \cref{eq:epsilon}, $U_0/R_0 = V_0/\lambda$. We simplify notation and define $\bar{G}_{N+1}$ as:
\begin{equation}
\bar{G}_{N+1} =  - \left \langle \frac{\partial \bar{W}_N}{\partial \bar{t}} \right \rangle  -  \sum_{l=1}^{N} \sum_{p=0}^{N-l+1} \left \langle  \frac{\partial \bar{W}_{N+1-l}}{\partial \langle \bar{c}\rangle^p } \frac{\partial \bar{G}_l}{\partial \bar{x}} \right \rangle  -  \left \langle \bar{V}_x \frac{\partial \bar{W}_N}{\partial \bar{x}} \right \rangle 
-\left \langle \bar{V}_r \frac{\partial \bar{W}_N}{\partial \bar{r} } \right \rangle.   
\end{equation}   Thus, $G_{N+1}$ has the following form: 
\begin{align}
    G_{N+1}  & = \left( \frac{R_0^{2} V_0}{\lambda D}  \right)^{N} \frac{V_0c_0}{\lambda}  \bar{G}_{N+1}.  
\label{induction step done for G n+1} 
\end{align}  
We use \cref{induction step done for G n+1} and write \cref{general eq for w n+1 as in robers} in non-dimensional variables. We obtain the following: 
\begin{equation} 
\frac{D}{R_0^2}  \left( \frac{\partial^2 W_{N+1}}{\partial \bar{r}^2} + \frac{1}{\bar{r}} \frac{\partial W_{N+1} }{\partial \bar{r} }\right) = \left( \frac{R_0^{2} V_0}{\lambda D}  \right)^{N} \frac{V_0c_0}{\lambda} \left(  \bar{H}_{N+1} \right). \label{eq:solving for wn+1 }  
\end{equation}
where $\bar{H}_{N+1}$ is a function in non-dimensional variables given by:
\[ \bar{H}_{N+1} = \frac{\partial \bar{W}_N}{\partial \bar{t}}  + \bar{G}_{N+1} + \sum_{l=1}^{N} \sum_{p=0}^{N-l+1}   \frac{\partial \bar{W}_{N+1-l}}{\partial \langle \bar{c}\rangle^p } \frac{\partial \bar{G}_l}{\partial \bar{x}}  +  \bar{V}_x \frac{\partial \bar{W}_N}{\partial \bar{x}} 
+ \bar{V}_r \frac{\partial \bar{W}_N}{\partial \bar{r} }. \] 
Then, we multiply \cref{eq:solving for wn+1 } by $\bar{r}$, integrate with respect to $\bar{r}$ and enforce boundedness of $\partial W_{N+1} / \partial r$ at $r =0$:
$$ \bar{r} \frac{\partial W_{N+1}}{\partial \bar{r}} =  \left( \frac{R_0^{2} V_0}{\lambda D}\right)^{N+1} c_0 \int_0^{\bar{r}} \bar{H}_{N+1}(s)sds .$$ 
We note that the impermeability condition \cref{eq:impermeability condition for W_i}  is satisfied due to the value of $\bar{G}_{N+1}$ and the definition of $\bar{H}_{N+1}$. We solve for $W_{N+1}$:
\begin{equation}
    W_{N+1} = \left(\frac{R_0^{2} V_0}{\lambda D}\right)^{N+1} c_0\left( \int_0^{\bar{r}} \frac{1}{z} \int_0^{z} \bar{H}_{N+1}(s)sdsdz + K \right) := \left(\frac{R_0^{2} V_0}{\lambda D}\right)^{N+1} c_0 \bar{W}_{N+1} .
\end{equation}
where the constant $K$ is chosen to ensure that \cref{eq:condition on W_i} is satisfied. Thus, we have shown that \cref{eq:nondim W} and \cref{eq:nondim G} hold for $n = N+1$. 
Writing \cref{eq:expression for the equation on the centre manifold} in non-dimensional form and using \cref{eq:nondim W} and \cref{eq:nondim G} yield the following. 
\begin{align}
\frac{V_0 c_0}{\lambda } \frac{\partial \langle \bar{c} \rangle}{\partial \bar{t} } = \sum_{n=1}^{\infty} \left( \frac{R_0^{2} V_0}{\lambda D} \right)^{n-1} \frac{V_0c_0}{\lambda}  \bar{G}_n.
\end{align}
Thus, in non-dimensional variables \cref{eq:nondim W} reads
\begin{equation}
 \frac{\partial \langle \bar{c} \rangle}{\partial \bar{t}  } = \sum_{n=1}^{\infty} \left( \frac{R_0^{2} V_0  }{\lambda D} \right)^{n-1} \bar{G}_n. 
\end{equation}
We define $\epsilon := (R_0^2 V_0)/(\lambda D)$. Using assumption \cref{eq:assumption to get convergence} and neglecting $\mathcal{O}(\epsilon^n), n \geq 2$ terms, we obtain 
\begin{equation} 
    \frac{\partial \langle \bar{c} \rangle}{\partial \bar{t}  }  = \bar{G_1} + \left( \frac{R_0^{2} V_0  }{\lambda D} \right)\bar{G}_2.
    \label{eq:nondim after neglecting}
\end{equation}
Rewriting \cref{eq:nondim after neglecting} in dimensional variables, we obtain the following. 
\[
 \frac{\partial \langle c \rangle}{\partial t }   = G_1 + G_2.
\]
Similarly, we write \cref{eq:expression for the equation on the centre manifold} in non-dimensional variables and neglect $\mathcal{O}(\epsilon^n), n\geq 2$ terms. We conclude that 
\[
c_0 \bar{c}   = c_0 \bar{W}_0 + \frac{R_0^2V_0}{\lambda D} c_0\bar{W}_1,
\]
which implies that $c = W_0+W_1$.
\end{proof}

\bibliographystyle{siamplain} 
\bibliography{references}

\begin{thebibliography}{10}

\bibitem{aris1956dispersion}
{\sc R.~Aris}, {\em On the dispersion of a solute in a fluid flowing through a
  tube}, {Proceedings of the Royal Society of London. Series A. Mathematical
  and Physical Sciences}, 235 (1956), p.~67 77.

\bibitem{azer2005taylor}
{\sc K.~Azer}, {\em Taylor diffusion in time dependent flow}, {International
  Journal of Heat and Mass Transfer}, 48 (2005), p.~2735 2740.

\bibitem{azer2007one}
{\sc K.~Azer and C.~S. Peskin}, {\em A one dimensional model of blood flow in
  arteries with friction and convection based on the womersley velocity
  profile}, {Cardiovascular Engineering}, 7 (2007), p.~51 73.

\bibitem{barnard1966theory}
{\sc A.~Barnard, W.~Hunt, W.~Timlake, and E.~Varley}, {\em A theory of fluid
  flow in compliant tubes}, {Biophysical Journal}, 6 (1966), p.~717 724.

\bibitem{boileau2015benchmark}
{\sc E.~Boileau, P.~Nithiarasu, P.~J. Blanco, L.~O. M{\"u}ller, F.~E. Fossan,
  L.~R. Hellevik, W.~P. Donders, W.~Huberts, M.~Willemet, and J.~Alastruey},
  {\em A benchmark study of numerical schemes for one-dimensional arterial
  blood flow modelling}, International journal for numerical methods in
  biomedical engineering, 31 (2015), p.~e02732.

\bibitem{vcanic2003mathematical}
{\sc S.~{\v{C}}ani{\'c} and E.~H. Kim}, {\em Mathematical analysis of the
  quasilinear effects in a hyperbolic model blood flow through compliant axi
  symmetric vessels}, {Mathematical Methods in the Applied Sciences}, 26
  (2003), p.~1161 1186.

\bibitem{carr1983application}
{\sc J.~Carr and R.~G. Muncaster}, {\em The application of centre manifolds to
  amplitude expansions. ii. infinite dimensional problems}, Journal of
  differential equations, 50 (1983), pp.~280--288.

\bibitem{cockburn1989tvb}
{\sc B.~Cockburn and C.~W. Shu}, {\em {TVB} {R}unge {K}utta local projection
  discontinuous {G}alerkin finite element method for conservation laws. ii.
  general framework}, Mathematics of {C}omputation, 52 (1989), p.~411 435.

\bibitem{coullet1983amplitude}
{\sc P.~Coullet and E.~A. Spiegel}, {\em Amplitude equations for systems with
  competing instabilities}, SIAM Journal on Applied Mathematics, 43 (1983),
  pp.~776--821.

\bibitem{d2007multiscale}
{\sc C.~D'Angelo}, {\em Multiscale modelling of metabolism and transport
  phenomena in living tissues}, tech. report, EPFL, 2007.

\bibitem{marbach2019active}
{\sc S.~Marbach and K.~Alim}, {\em Active control of dispersion within a
  channel with flow and pulsating walls}, Physical Review Fluids, 4 (2019),
  p.~114202.

\bibitem{mercer1990centre}
{\sc G.~Mercer and A.~Roberts}, {\em A centre manifold description of
  contaminant dispersion in channels with varying flow properties}, {SIAM
  Journal on Applied Mathematics}, 50 (1990), p.~1547 1565.

\bibitem{mohammed2014modelling}
{\sc F.~Mohammed, D.~Ngo-Cong, D.~Strunin, N.~Mai-Duy, and T.~Tran-Cong}, {\em
  Modelling dispersion in laminar and turbulent flows in an open channel based
  on centre manifolds using 1d-irbfn method}, Applied Mathematical Modelling,
  38 (2014), pp.~3672--3691.

\bibitem{mynard2015one}
{\sc J.~P. Mynard and J.~J. Smolich}, {\em One-dimensional haemodynamic
  modeling and wave dynamics in the entire adult circulation}, Annals of
  biomedical engineering, 43 (2015), pp.~1443--1460.

\bibitem{olufsen2000numerical}
{\sc M.~S. Olufsen, C.~S. Peskin, W.~Y. Kim, E.~M. Pedersen, A.~Nadim, and
  J.~Larsen}, {\em Numerical simulation and experimental validation of blood
  flow in arteries with structured tree outflow conditions}, {Annals of
  Biomedical Engineering}, 28 (2000), p.~1281 1299.

\bibitem{puelz2017computational}
{\sc C.~Puelz, S.~Acosta, B.~Rivi{\`e}re, D.~J. Penny, K.~M. Brady, and C.~G.
  Rusin}, {\em A computational study of the fontan circulation with
  fenestration or hepatic vein exclusion}, {Computers in Biology and Medicine},
  89 (2017), p.~405 418.

\bibitem{puelz2017comparison}
{\sc C.~Puelz, S.~{\v{C}}ani{\'c}, B.~Rivi{\`e}re, and C.~G. Rusin}, {\em
  Comparison of reduced models for blood flow using {R}unge {K}utta
  discontinuous {G}alerkin methods}, {Applied Numerical Mathematics}, 115
  (2017), p.~114 141.

\bibitem{roberts1988application}
{\sc A.~Roberts}, {\em The application of centre-manifold theory to the
  evolution of system which vary slowly in space}, The ANZIAM Journal, 29
  (1988), pp.~480--500.

\bibitem{roberts1997low}
{\sc A.~Roberts}, {\em Low-dimensional modelling of dynamical systems}, arXiv
  preprint chao-dyn/9705010,  (1997).

\bibitem{taylor1953dispersion}
{\sc G.~I. Taylor}, {\em Dispersion of soluble matter in solvent flowing slowly
  through a tube}, Proc. R. Soc. Lond. A, 219 (1953), p.~186 203.

\bibitem{taylor1954conditions}
{\sc G.~I. Taylor}, {\em Conditions under which dispersion of a solute in a
  stream of solvent can be used to measure molecular diffusion}, Proc. R. Soc.
  Lond. A, 225 (1954), p.~473 477.

\bibitem{wiggins2003introduction}
{\sc S.~Wiggins}, {\em Introduction to applied nonlinear dynamical systems and
  chaos}, vol.~2, Springer Science \& Business Media, 2003.

\bibitem{young1991shear}
{\sc W.~a. Young and S.~Jones}, {\em Shear dispersion}, Physics of Fluids A:
  Fluid Dynamics, 3 (1991), pp.~1087--1101.

\end{thebibliography}
\end{document}